\documentclass[12pt,letterpaper]{article}

\newif\ifpdf  
 
\ifx\pdfoutput\undefined
  \pdffalse
\else
  \pdfoutput=1
  \pdftrue
\fi

\ifpdf
  \usepackage{graphicx,here,theorem,amssymb,amsmath,url,thumbpdf}
  \DeclareGraphicsExtensions{.pdf,.jpg,.jpeg}
  \usepackage[pdftex,pdfpagemode=None,
     pdfstartview={XYZ},]{hyperref}  
\else
  \usepackage{graphicx,here,theorem,amssymb,amsmath,url}
  \DeclareGraphicsExtensions{.eps,.ps,.eps.gz,.ps.gz}
  \newcommand{\href}[2]{{#2}}
\fi

\newcommand{\st}{\ensuremath{s_\mathrm{true}}}
\newcommand{\et}{\ensuremath{\epsilon_\mathrm{true}}}

\newcommand{\penn}{$^\mathrm{a}$}
\newcommand{\brandeis}{$^\mathrm{b}$}
\newcommand{\rutgers}{$^\mathrm{c}$}
\newcommand{\rockefeller}{$^\mathrm{d}$}
\newcommand{\oxford}{$^\mathrm{e}$}
\newcommand{\pisa}{$^\mathrm{f}$}
\newcommand{\toronto}{$^\mathrm{g}$}

\title{{\hfill\small CDF/MEMO/STATISTICS/PUBLIC/7117}\\\hfill\\
\bf\large Interval estimation in the presence of nuisance parameters.
1.~Bayesian approach.}

\author{
Joel Heinrich\penn,
Craig Blocker\brandeis,
John Conway\rutgers,
Luc Demortier\rockefeller, \and
Louis Lyons\oxford,
Giovanni Punzi\pisa,
Pekka K.~Sinervo\toronto \and
\scriptsize\it\penn%
University of Pennsylvania, Philadelphia, Pennsylvania 19104 \and
\scriptsize\it\brandeis%
Brandeis University, Waltham, Massachusetts 02254 \and
\scriptsize\it \phantom{xxxxx} \rutgers%
Rutgers University, Piscataway, New Jersey 08855 \phantom{xxxxx} \and
\scriptsize\it\rockefeller%
Rockefeller University, New York, New York 10021 \and
\scriptsize\it\oxford%
University of Oxford, Oxford OX1 3RH, United Kingdom \and
\scriptsize\it\pisa%
Istituto Nazionale di Fisica Nucleare,
University and Scuola Normale Superiore of Pisa, I-56100 Pisa, Italy \and
\scriptsize\it\toronto%
University of Toronto, Toronto M5S 1A7, Canada}
 
\date{September 27, 2004}  

\begin{document}

\maketitle

\begin{abstract}
We address the common problem of calculating intervals in the presence
of systematic uncertainties.  We aim to investigate several
approaches, but here describe just a Bayesian technique for setting
upper limits.  The particular example we study is that of inferring
the rate of a Poisson process when there are uncertainties on the
acceptance and the background.  Limit calculating software associated
with this work is available in the form of C~functions.

\end{abstract}

\section{The problem}
\label{problem}
A very common statistical procedure is obtaining a confidence interval
for a physics parameter of interest,
when there are uncertainties in quantities such as the acceptance of the
detector and/or the analysis procedure, the beam intensity, and the
estimated background. These are known in statistics as nuisance
parameters, or in Particle Physics as sources of systematic
uncertainty. We assume that estimates of these quantities are
available from subsidiary measurements.\footnote{
There are other possibilities. Thus it may be that all that is known is
that a nuisance parameter is contained within a certain range:
$\mu_\mathrm{l}\le\mu\le\mu_\mathrm{u}$;
that is not enough information for a Bayesian
approach. Alternatively the data relevant for the physics and nuisance
parameters could be bound up in the main measurement, and not require a
subsidiary one.}
A variant of
this procedure which is particularly relevant for Particle Physics is
the extraction of an upper limit on the rate of some hypothesized
process or on a physical parameter, again with systematic uncertainties.

To specify the problem in more detail, we assume that we are
performing a counting experiment in which we observe $n$ counts, and
that the acceptance has been estimated as $\epsilon_0 \pm
\sigma_\epsilon$ and the background as $b_0 \pm \sigma_b$.  For a
signal rate $s$, $n$ is Poisson distributed with mean $s\epsilon + b$.
Here $\epsilon$ contains factors like the intensity of the accelerator
beam(s), the running time, and various efficiencies. It is constrained
to be non-negative, but can be larger than unity.

We aim to study and compare different approaches
for determining confidence intervals for
this problem. In
general we are interested in pathologies in these areas:
\begin{itemize}

\item
{\bf  Coverage.}
This is a measure of how often the limits that we deduce would
in fact include the true value of the parameter. This requires consideration of an ensemble 
of experiments like the one we actually performed, and hence is an essentially 
frequentist concept. Nevertheless, it can be applied to a Bayesian technique.

Coverage is a property of the technique, and not of the particular
limit deduced from a given measurement. It can, however, be a function
of the true value of the parameter, which is in general unknown in a
real measurement.

Undercoverage (i.e.\ the probability of containing 
the true value is less than the stated confidence level) is regarded by frequentists as a
serious defect.
Usually coverage is required for all possible values of the physical
parameter.\footnote{The argument is that the parameter is unknown, and
so we wish to have coverage, whatever its value. This ensures that, if
we repeat our specific experiment many times, we should include the
true value within our confidence ranges in (at least) the stated
fraction of cases. This argument may, however, be over-cautious. The
location of the dips in a coverage plot like that of
Fig.~\ref{fig:Bayes} occur at values which are not fixed in $s$, but
which depend on the details of our experiment (such as the values of
$\epsilon$ and $b$). These details vary from experiment to
experiment. Thus we could achieve `no undercoverage for the ensemble
of experiments measuring the parameter $s$', even if the individual
coverage plots did fall below the nominal coverage occasionally.  Thus
in some sense `average coverage' would be sufficient (see
for example reference \cite{interplay}), although it is
hard to quantify the exact meaning of `average'.  It should be stated
that this is not the accepted position of most High Energy Physics
frequentists.}
In contrast, overcoverage is permissible, but the larger intervals result
in less stringent tests of models that predict the value of the parameter. For 
measurements involving quantised data (e.g.\ Poisson counting), most
methods have coverage which varies with the true value of the parameter of interest,
and hence if undercoverage is to be avoided, overcoverage is inevitable. 

Frequentist methods by construction will not undercover for any values
of the parameters. This is not guaranteed for other approaches. For
example, even though the Bayesian intervals shown here do not
undercover, in other problems Bayesian 95\% credible intervals could
even have zero coverage for some values of the parameter of
interest.\cite{zeroc}
It should also be remarked that, although coverage is a very important
property for frequentists, on its own exact coverage does not
guarantee that intervals have desirable properties (for many examples,
see Refs. \cite{clifford} and \cite{strong}).



\item
{\bf Interval length.}
This is sometimes used as a criterion of accuracy of
intervals, in the sense that shorter intervals have less probability of
covering false values of the parameter of interest.  However, one should
keep in mind that short intervals are only desirable if they contain the
true value of the parameter.  Thus empty intervals, which do occur in
some frequentist constructions, are generally undesirable, even when their
construction formally enjoys frequentist coverage.

Intervals that fill the entire physically allowed range of the
parameter of interest may also occur in some situations.  Examples of
this behavior are given in \cite{clifford} and \cite{zech}.  An
experimenter who requests a 68\% confidence interval, but receives what
appears to be a 100\% confidence interval instead, may not be satisfied
with the explanation that he is performing a useful service in helping
to keep the coverage probability---averaged over his measurement and
his competitor's measurements---from dropping below 68\%.


\item
{\bf Bayesian credibility.}
In some situations it may be relevant to calculate the Bayesian
credibility of an interval, even when the latter was constructed by
frequentist methods.  This would of course require one to choose a
prior for all the unknown parameters.  The question is one of
plausibility: given the type of measurement we are making, the
resolution of the apparatus, etc., how likely is it that the true
value of the parameter we are interested in lies in the calculated
interval?  Does this
differ dramatically from 
the nominal coverage probability of the interval?
In fact for different values of the observable(s), frequentist ranges
are very likely to have different credibilities. Some examples
of this behavior are noted in Ref.~\cite{karlen}.

When calculating the Bayesian credibility of frequentist intervals,
``uninformative'' priors appear advisable.  Note that severe interval
length pathologies will automatically produce a large inconsistency
between the nominal coverage and the Bayesian credibility of an
interval.  Except in a handful of very special cases, it is not
possible to construct an interval scheme that has simultaneously
constant Bayesian credibility and constant frequentist coverage, even
if one has total freedom in choosing the prior(s).  Although it is not
at all clear exactly how large a level of disagreement is
pathological, nevertheless it may be instructive to know how severely
an interval scheme deviates from constant Bayesian credibility (and
how sensitive this is to the choice of prior).

\item
{\bf  Bias.}
In the context of interval selection, this means having a larger
coverage $B(s',\st)$ for an incorrect value $s'$ of the parameter than
for the true value \st. This requires plots of coverage versus $s'$
for different values of \st. For upper limits, $B(s_1',\st) \ge
B(s_2',\st)$ if $s_1'$ is less than $s_2'$, so methods are necessarily
biassed for low $s'$.
Bias thus is not very interesting for upper limits.
It will be discussed in
later notes dealing with two-sided intervals.

\item
{\bf Transformation under reparametrisation.}
Intervals that are not transformation-respecting can be problematic. For
example, it is possible for the predicted value of the lifetime of a
particle to be contained within the 90\% interval determined from the
data, but for the corresponding predicted value of the decay rate (equal
to the reciprocal of the predicted lifetime) to be outside the 90\%
interval when the data is analysed by the same procedure, but in terms
of decay rate. This would result in unwanted ambiguities about the
compatibility of the data with the prediction.

\item
{\bf Unphysical ranges.}
The question here is whether the interval
construction procedure can be made to respect the physical boundaries
of the problem.
For example, branching fractions should be in the range zero to one,
masses should not be negative, etc. Statements about the {\em true}
value of a parameter should respect any physical bounds.  In contrast,
some methods give {\em estimates} of parameters which can themselves
be unphysical, or which include unphysical values when the errors are
taken into account. We do not recommend truncating ranges of {\em
estimates} of parameters to obey such bounds. Thus the fact that a
branching fraction is estimated as $1.1\pm0.2$ conveys more
information about the experimental result than does the statement that
it lies in the range 0.9 to 1.

\item
{\bf Behavior with respect to nuisance parameter.}
We would normally expect that the limits on a physical parameter would
tighten as the uncertainty on a nuisance parameter decreases; and that
as this uncertainty tends to zero, the limits should agree with those
obtained under the assumption that the ``nuisance parameter'' was
exactly known. (Otherwise we could sometimes obtain a tighter limit
simply by pretending that we knew less about the nuisance parameter
than in fact is the case.) These desiderata are not always satisfied
by non-Bayesian methods (see \cite{ch} and \cite{feldman}).

\end{itemize}

Although we are ultimately interested in comparing different
approaches to this problem, in this note we investigate a Bayesian
technique for determining upper limits. Our purpose is to spell out in
some detail how this approach is used, and to discuss some of the
properties of the resulting limits in this specific example. We
believe that, for variants of this problem (e.g.\ different choice of
prior for $s$; alternative assumptions about the information on the
nuisance parameters; etc.), the reader could readily adapt the
techniques described here (and the associated software) to their
particular situation.

We will report on two-sided intervals and also compare with
other methods (e.g.\ Cousins--Highland, pure frequentist,
profiled frequentist) in later notes.

\section{Reminder of Bayesian approach}
\label{Bayes}

Before dealing with the problem of extracting and studying the limits on $s$ as deduced
from observing $n$ events from a
Poisson distribution with mean $s\epsilon + b$
in the presence of an uncertainty on $\epsilon$, we recall the way 
the Bayesian approach works for the simpler problem of a counting experiment with no 
background and with $\epsilon$ exactly known. Then $n$ is Poisson
distributed with mean $s\epsilon$,
and Bayes' Theorem\footnote{We follow the common convention whereby
lower case $\pi$'s denote prior p.d.f.'s, lower case $p$'s denote
other p.d.f.'s, upper case $\Pi$'s denote prior probabilities, and
upper case $P$'s denote other
probabilities. Equation~(\ref{eqn:BayesTh}) is true for probabilities,
p.d.f.'s, or mixtures depending on whether $B$ and/or $C$ are
discrete or continuous variables.}
\begin{equation}
P(B|C) = P(C|B)P(B)/P(C)
\label{eqn:BayesTh}
\end{equation}
gives
\begin{equation}
p(s|n) = \frac{P(n|s)  \pi(s)}{\int P(n|s)  \pi(s) \ ds}                            
\label{eqn:P(s|n)}
\end{equation}
where $\pi(s)$ is the prior probability density for $s$;
$p(s|n)$ is the posterior probability density function (p.d.f.)\ for $s$, given 
the observed $n$; and
$P(n|s)$ is the probability of observing $n$, given $s$.

We assume a constant prior for $s$,\footnote{This is an assumption, not a necessity, and 
is in some ways unsatisfactory. (It is implausible, cannot be normalised, and 
creates divergences for the posterior if used with a (truncated) Gaussian prior 
for the acceptance $\epsilon$.)} and that $P(n|s)$ is given by the Poisson 
\begin{equation}
P(n|s) = e^{-s\epsilon}(s\epsilon)^n/n!
\label{eqn:Poisson}
\end{equation}
Then\footnote{It turns out that the sum over $n$ of the discrete distribution 
(\ref{eqn:Poisson}) and the integral over $s$ of the continuous distribution 
(\ref{eqn:Poisson2}) are both equal to unity.
This means that the probability $P(n|s)$ and the probability density
$p(s|n)$ are correctly normalised.}
\begin{equation}
p(s|n) = \epsilon e^{-s\epsilon}(s\epsilon)^n/n!
\label{eqn:Poisson2}
\end{equation}
The limit is now obtained by integrating this posterior p.d.f.\ for $s$ until we
achieve the required fraction $\beta$ of the total integral from zero to infinity. If 
$\beta$ is $90\%$, the upper limit $s_\mathrm{u}$ is given by
\begin{equation}
\int^{s_\mathrm{u}}_0\!\! p(s|n)\ ds =0.9
\label{eqn:0.9}
\end{equation}
$\beta$ is termed the credible or Bayesian confidence level for the limit. 

For different observed $n$, the upper limits are shown in the last two
columns of Table~\ref{ltablex}, for $b = 0$ and for $b = 3$
respectively.
The Gaussian
approximation for the case $b = 0$, $n=20$, would yield
$s_\mathrm{u}\simeq20+1.28\sqrt{20}\simeq25.7$, which is roughly
comparable to the corresponding $s_\mathrm{u}=27.0451$ of the Table.
For $b = 0$, it coincidentally turns out that, for this
particular example, the Bayesian upper limits are identical with those
obtained in a frequentist calculation with the Neyman construction and
a simple ordering rule (see later note on the frequentist approach to
this problem). In general this is not so.
Other priors sometimes used for $s$ are $1/\sqrt s$ \cite{roots} or
$1/s$ \cite{ref:Jeffreys}. Having a prior peaked at smaller values of
$s$ in general results in tighter limits for a given observed $n$.

If the whole procedure is now repeated with a background $b$ and a
flat prior, the upper limits not surprisingly decrease for increasing
$b$ at fixed $n$ (except for the case $n = 0$ where the limits can
trivially be seen to be independent of $b$). This is not inconsistent
with the fact that the mean limit for a series of measurements
increases with $b$, i.e.\ experiments with larger expected backgrounds
have poorer sensitivity.

\begin{table}
\begin{center}
\scriptsize
\begin{tabular}{@{}r|rrrrrrrrr|rr@{}}
\hline
&\multicolumn{9}{c|}{$\epsilon=1.0\pm0.1$}&
\multicolumn{2}{c}{$\epsilon=1\pm0$}\\
$n$&\multicolumn{1}{c}{$b=0$}&
\multicolumn{1}{c}{1}&
\multicolumn{1}{c}{2}&
\multicolumn{1}{c}{3}&
\multicolumn{1}{c}{4}&
\multicolumn{1}{c}{5}&
\multicolumn{1}{c}{6}&
\multicolumn{1}{c}{7}&
\multicolumn{1}{c|}{8}&
\multicolumn{1}{c}{$b=0$}&
\multicolumn{1}{c}{3}\\
\hline
0&2.3531&2.3531&2.3531&2.3531&2.3531&2.3531&2.3531&2.3531&2.3531&2.3026&2.3026\\
1&3.9868&3.3470&3.0620&2.9019&2.8000&2.7297&2.6783&2.6391&2.6083&3.8897&2.8389\\
2&5.4669&4.5520&3.9676&3.6026&3.3623&3.1953&3.0736&2.9816&2.9099&5.3223&3.5228\\
3&6.8745&5.8618&5.0463&4.4644&4.0571&3.7666&3.5534&3.3922&3.2671&6.6808&4.3624\\
4&8.2380&7.1964&6.2451&5.4751&4.8914&4.4569&4.1313&3.8832&3.6904&7.9936&5.3447\\
5&9.5714&8.5213&7.5063&6.6022&5.8579&5.2719&4.8180&4.4660&4.1904&9.2747&6.4371\\
6&10.8826&9.8288&8.7885&7.8047&6.9344&6.2066&5.6184&5.1499&4.7772&10.5321&7.5993\\
7&12.1766&11.1203&10.0703&9.0460&8.0904&7.2450&6.5289&5.9387&5.4586&11.7709&8.7958\\
8&13.4570&12.3984&11.3441&10.3014&9.2952&8.3635&7.5374&6.8300&6.2380&12.9947&10.0030\\
9&14.7261&13.6655&12.6085&11.5575&10.5247&9.5365&8.6250&7.8142&7.1136&14.2060&11.2085\\
10&15.9858&14.9233&13.8641&12.8090&11.7630&10.7415&9.7701&8.8758&8.0775&15.4066&12.4073\\
11&17.2375&16.1732&15.1121&14.0542&13.0017&11.9621&10.9525&9.9966&9.1170&16.5981&13.5983\\
12&18.4823&17.4163&16.3533&15.2934&14.2371&13.1881&12.1560&11.1582&10.2162&17.7816&14.7816\\
13&19.7210&18.6535&17.5887&16.5269&15.4682&14.4139&13.3692&12.3452&11.3588&18.9580&15.9580\\
14&20.9545&19.8854&18.8191&17.7554&16.6946&15.6373&14.5856&13.5459&12.5302&20.1280&17.1280\\
15&22.1832&21.1127&20.0448&18.9795&17.9169&16.8572&15.8014&14.7528&13.7187&21.2924&18.2924\\
16&23.4078&22.3359&21.2665&20.1996&19.1353&18.0737&17.0151&15.9612&14.9161&22.4516&19.4516\\
17&24.6286&23.5553&22.4845&21.4161&20.3502&19.2868&18.2261&17.1689&16.1172&23.6061&20.6061\\
18&25.8459&24.7714&23.6992&22.6294&21.5619&20.4969&19.4344&18.3747&17.3189&24.7563&21.7563\\
19&27.0601&25.9844&24.9109&23.8397&22.7708&21.7042&20.6400&19.5784&18.5198&25.9025&22.9025\\
20&28.2715&27.1946&26.1199&25.0474&23.9770&22.9090&21.8432&20.7799&19.7191&27.0451&24.0451\\
\hline
\end{tabular}
\end{center}
\caption{Upper 90\% limits for $n$ observed events with $b$ background
and $\epsilon=1.0\pm0.1$ ($\kappa=100$ and $m=99$,
as defined in section~\ref{submeas}). Also shown
are limits for $b=0$ and $b=3$ with fixed $\epsilon=1$.
\label{ltablex}}
\end{table}


\subsection{Coverage}

Next we can investigate the frequentist coverage $C(\st)$\footnote 
{This is the coverage at $s = \st$ when the Poisson variable is generated with 
$s = \st$. This differs from $B(s',\st)$ where the coverage is checked at 
$s = s'$ when the generation value is \st.} of this Bayesian approach. 
That is, we can ask what the probability is, for a given value of \st, of our upper 
limit being larger than \st, and hence being consistent with it. This is equivalent 
to adding up the Poisson probabilities 
of eqn.~(\ref{eqn:Poisson}) for those values of $n$ for which $s_\mathrm{u}(n) \ge \st$ i.e.
\begin{equation}
C(\st) = \sum_{\mathrm{relevant}\ n}\!\!\!e^{-\st\epsilon}(\st\epsilon)^n / n!
\label{csum}
\end{equation}
As \st\ increases through any of the values of $s_\mathrm{u}$ of
the last two columns of
Table~\ref{ltablex}, the coverage drops sharply.
For example, for the case of zero background and efficiency known to
be unity, the 90\% Bayesian upper limits will include $\st=3.8896$ for
$n = 1$ or larger. But $\st=3.8898$ is no longer below the upper limit
for $n = 1$. Thus one term drops out of the summation of
eqn.~(\ref{csum}) for the calculation of the coverage at $\st=3.8898$,
while the remaining terms change but little for the small change in
\st; this produces the abrupt fall in coverage. The coverage is
plotted in Fig.~\ref{fig:Bayes}, where the drop at $\st=3.8897$ can be
seen.


{\footnotesize
\begin{quotation}
The calculation of $C(\st)$ can be done as follows:
The identity
\begin{equation}
f'(x)=
e^{-x}\left[\sum_{k=0}^{n-1}{x^k\over k!}-\sum_{k=0}^n{x^k\over k!}\right]=
-e^{-x}{x^n\over n!}
\qquad\mathrm{for}\qquad f(x)=e^{-x}\sum_{k=0}^n{x^k\over k!}
\end{equation}
allows us to write (integrating $-f'(x)$)
\begin{equation}
\int^{\st}_0\!\!\!\! p(s|n)\,ds =
\int^{\st\epsilon}_0\!\!\!\!e^{-x}{x^n\over n!}dx =
1-e^{-\st\epsilon}\sum_{k=0}^n{(\st\epsilon)^k\over k!}
\end{equation}
From this, it follows that
``relevant $n$'' is equivalent to any one of these
inequalities:
\begin{equation}
s_\mathrm{u}(n) \ge \st\Leftrightarrow
\int^{s_\mathrm{u}}_0\!\! p(s|n)\,ds\ge\int^{\st}_0\!\!\!\! p(s|n)\,ds\Leftrightarrow
\beta\ge1-e^{-\st\epsilon}\sum_{k=0}^n{(\st\epsilon)^k\over k!}
\end{equation}
and our expression for the coverage becomes
\begin{equation}
C(\st) = 1 - {\sum_{n=0}}'e^{-\st\epsilon}{(\st\epsilon)^n \over n!}
\end{equation}
where ${\sum}'$ means ``sum until the next term would cause the sum to
exceed $1-\beta$''. This result proves that $C(\st)\ge\beta$
for all values of \st\ in this simple example.

\end{quotation}
}


It is seen that the coverage starts at $100\%$ for small \st. This is because 
even for $n = 0$  the Bayesian upper limit will include \st, and this is even more
so for larger $n$.

Bayesian methods can be shown to achieve average coverage. By this we
mean that when the coverage is averaged over the parameter $s$,
weighted by the prior in $s$, the result will agree with the nominal
value $\beta$, i.e.
\begin{equation}
\frac{\int C(s)\ \pi(s)\ ds}{\int \pi(s)\ ds} = \beta
\end{equation} 
A proof of this theorem is given in the second appendix, section~\ref{ac} of this note.
 
For a constant prior, the region at large $s$ tends to dominate the
average, while in general we will be interested in the coverage at
small $s$. Thus the ``average coverage'' result is of academic rather
than practical interest, especially for the case of a flat prior.
Indeed it is possible to have a situation where the average coverage
is, say, $90\%$, while the coverage as a function of $s$ is always
larger than or equal to $90\%$.

\section{The actual problem}
Our actual problem differs from the simple case of Section \ref{Bayes}
in that\\ (a) we have a background $b$, assumed for the time being to
be accurately known; and (b) we have an acceptance $\epsilon$
estimated in a subsidiary experiment as $\epsilon_0 \pm
\sigma_\epsilon$.

What we are going to do is to use a multidimensional version of Bayes' Theorem
to express $p(s,\epsilon|n)$ in terms of $P(n|s,\epsilon)$ and the
priors for $s$ and $\epsilon$. The relationship is\footnote{
For the case where the probabilities have a frequency ratio
interpretation, this is seen
from the mathematical identities\\ \\
$P(X\ \mathrm{and}\ Y\ \mathrm{and}\ Z) = \frac{N(X\ \mathrm{and}\ Y\ \mathrm{and}\ Z)}{N(Z)}\frac{N(Z)}{N_\mathrm{tot}} =P(X,Y|Z)\ 
P(Z)$
and \\ \\
$P(X\ \mathrm{and}\ Y\ \mathrm{and}\ Z) = \frac{N(X\ \mathrm{and}\ Y\ \mathrm{and}\ Z)}{N(X\ \mathrm{and}\ Y)}\frac{N(X\ \mathrm{and}\ Y)}{N_\mathrm{tot}} =
P(Z|X,Y)\ P(X,Y)$.\\ \\
So with $X$, $Y$ and $Z$ identified with $s$, $\epsilon$ and $n$ respectively, and with the prior for
$s$ and $\epsilon$ factorising into two separate priors for $s$ and for $\epsilon$, we obtain
$p(s,\epsilon|n)\ P(n) = P(n|s,\epsilon)\ \pi(s)\ \pi(\epsilon)$.
}
\begin{equation}
p(s,\epsilon|n) = \frac{P(n|s,\epsilon)\pi(s)\pi(\epsilon)}{\int\!\!\int 
P(n|s,\epsilon)\pi(s)\pi(\epsilon)\ ds\ d\epsilon}
\label{eqn:extended}
\end{equation}

To obtain the posterior p.d.f.\ for $s$, we now integrate this over $\epsilon$: 
\begin{equation}
p(s|n) = \int_0^\infty\!\! p(s,\epsilon|n)\ d\epsilon,
\label{eqn:posterior}
\end{equation}
and finally we use this to set a limit on $s$ as in eqn.~(\ref{eqn:0.9}).

The coverage for this procedure needs to be calculated as a function of 
\st\ and \et.
The average coverage theorem of the previous section must be
generalized to
\begin{equation}
{\int\!\!\int C(s,\epsilon)\pi(s)\pi(\epsilon)\,ds\,d\epsilon
\over
\int\!\!\int \pi(s)\pi(\epsilon)\,ds\,d\epsilon}=\beta
\end{equation}

\subsection{Priors}
To implement the above procedure we need priors for $s$ and $\epsilon$. As in 
the simple example of Section \ref{Bayes}, for simplicity we assume 
that the prior for $s$ is constant. It will be interesting to look at 
the way the properties of this method change as other priors for $s$ are used.

We assume that the prior for $\epsilon$ is extracted from some
subsidiary measurement $\epsilon_0 \pm \sigma_\epsilon$. We do {\bf
not} assume that this implies that our belief about \et\ is
represented by a {\bf Gaussian} distribution centred on $\epsilon_0$,
as this would give trouble with the lower end of the Gaussian
extending to negative $\epsilon$.
Instead, we specify some particular form of the
subsidiary experiment that provides information about $\epsilon$, and
then assume that a Bayesian analysis of this yields a posterior
p.d.f.\ for $\epsilon$. Slightly confusingly, this posterior from the
subsidiary experiment is used as the prior for the application of
Bayes' Theorem to extract the limit on $s$ (see eqns.\
(\ref{eqn:extended}) and (\ref{eqn:posterior})).

\subsection{The subsidiary measurement}
\label{submeas}
\begin{figure}[p]
\begin{center}
\includegraphics[width=\textwidth]{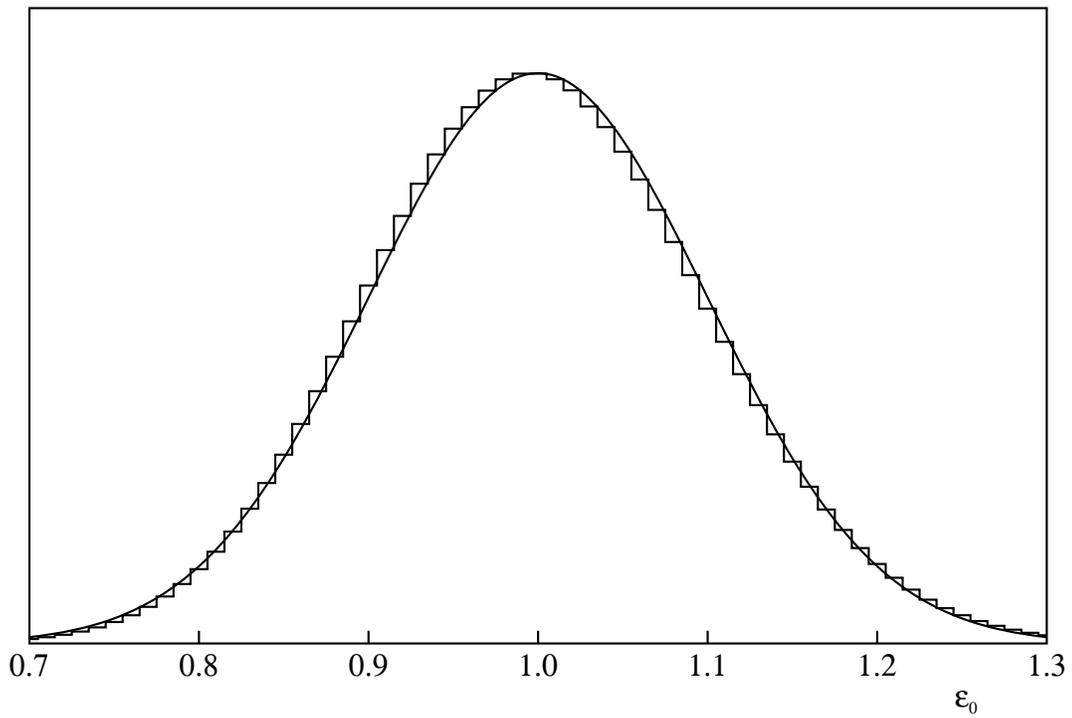}
\caption{
Comparison of our discrete probability for $\epsilon_0$ (shown as
a histogram, see eqn.~(\ref{eqn:pdf})) and Gaussian (continuous curve)
for the case $\epsilon=1.0\pm0.1$.}
\label{fig:comparison}
\end{center}
\end{figure}

Somewhat arbitrarily, we assume that, for a true acceptance \et, the 
probability  for the measured value $\epsilon_0$ in the subsidiary experiment
is given by a Poisson distribution 
\begin{equation}
P(\epsilon_0|\et) = e^{-\kappa\et}\kappa^m\et^m/m!
\label{eqn:pdf}
\end{equation}
where $\epsilon_0 = (m+1)/\kappa$, $\sigma_\epsilon^2 =
(m+1)/\kappa^2$ and $\kappa$ is a scaling constant\footnote{Here we define
$\epsilon_0$ and $\sigma^2_{\epsilon}$ as the mean and variance of the
posterior p.d.f.\ of eqn.~(\ref{eqn:posterior_e}).}. We interpret this
as the probability for $\epsilon_0$. This is discrete because the
observable $m$ is discrete, but the allowed values become closely
spaced for large $\kappa$. For small $\sigma_\epsilon/\epsilon$
(i.e.\ for large $m$), these probabilities approximate to a narrow Gaussian
(see Fig.~\ref{fig:comparison}).

\begin{figure}[p]
\begin{center}
\includegraphics[width=\textwidth]{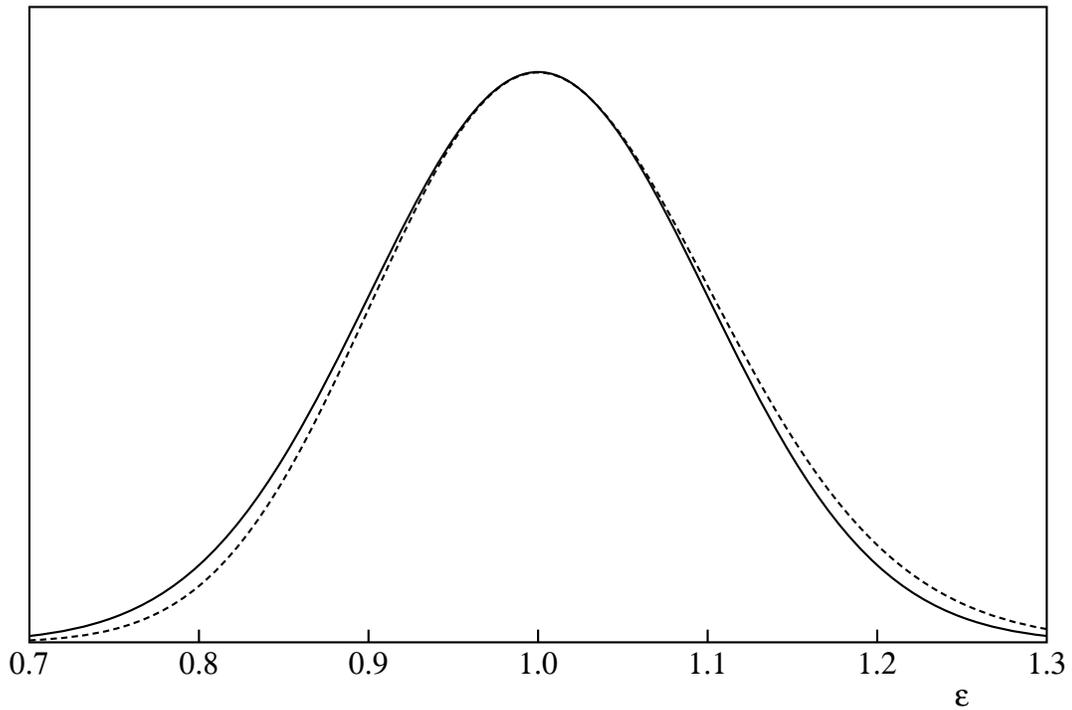}
\caption{Comparison of our likelihood
(dashed, see eqn.~(\ref{eqn:likelihood}))
and Gaussian (solid) for the case
$\epsilon=1.0\pm0.1$.}
\label{fig:comparison2}
\end{center}
\end{figure}

Given our choice of probability in eqn.~(\ref{eqn:pdf}),
the likelihood for the parameter $\epsilon$, given measured $\epsilon_0$,
is
\begin{equation}
\mathcal{L}(\epsilon|\epsilon_0) = e^{-\kappa\epsilon}\kappa^m\epsilon^m/m!
\label{eqn:likelihood}
\end{equation}
This is the same function of $\epsilon$ and $\epsilon_0$ as eqn.~(\ref{eqn:pdf}), but 
now $m$
is regarded as fixed, and $\epsilon$ is the variable. The likelihood is a continuous function of
$\epsilon$. It is compared with a Gaussian in Fig.~\ref{fig:comparison2}.

Finally in the Bayes approach, with the choice of a constant prior for $\epsilon$, the posterior
probability density for $\epsilon$
 after our subsidiary measurement is 
\begin{equation}
p(\epsilon|m) \propto \ e^{-\kappa\epsilon}\kappa^m\epsilon^m/m!
\label{eqn:posterior_e}
\end{equation}
which is obtained by multiplying the right-hand side of eqn.~(\ref{eqn:likelihood}) by unity.
This {\bf posterior} probability density for $\epsilon$ will be used as our {\bf prior} 
for $\epsilon$ in the next step of deducing the limit for $s$.

\section{Results}

The details of the necessary analytical calculations\footnote{This
example can be handled analytically. More complicated cases might require
numerical integration, which can be done via numerical quadrature or
Monte Carlo methods.}  are presented in the Appendix of this note. In
this section we investigate the behavior of the Bayesian limits in
this example, especially the shape of the frequentist coverage
probability as a function of
\st.

\subsection{Shape of the posterior}

The posterior p.d.f.\ for $s$ has the form
\begin{equation}
p(s|b,n)ds\propto
\left[\int_0^\infty
{e^{-(\epsilon s+b)}(\epsilon s+b)^n\over n!}
{\kappa  (\kappa\epsilon)^{m}  e^{-\kappa\epsilon} \over\Gamma(m+1)}d\epsilon
\right]1\,ds
\end{equation}
where the likelihood, the prior for $\epsilon$, the (constant) prior
for $s$, and the marginalization integral over $\epsilon$ are all
prominently displayed.

The posterior probability density for $s$ gives the complete summary
of the outcome of the measurement in the Bayesian approach.  It is
therefore important to understand its shape before proceeding to use
it to compute a limit (or extract a central value and error-bars).

\begin{figure}[p]
\begin{center}
\includegraphics[width=3.15in]{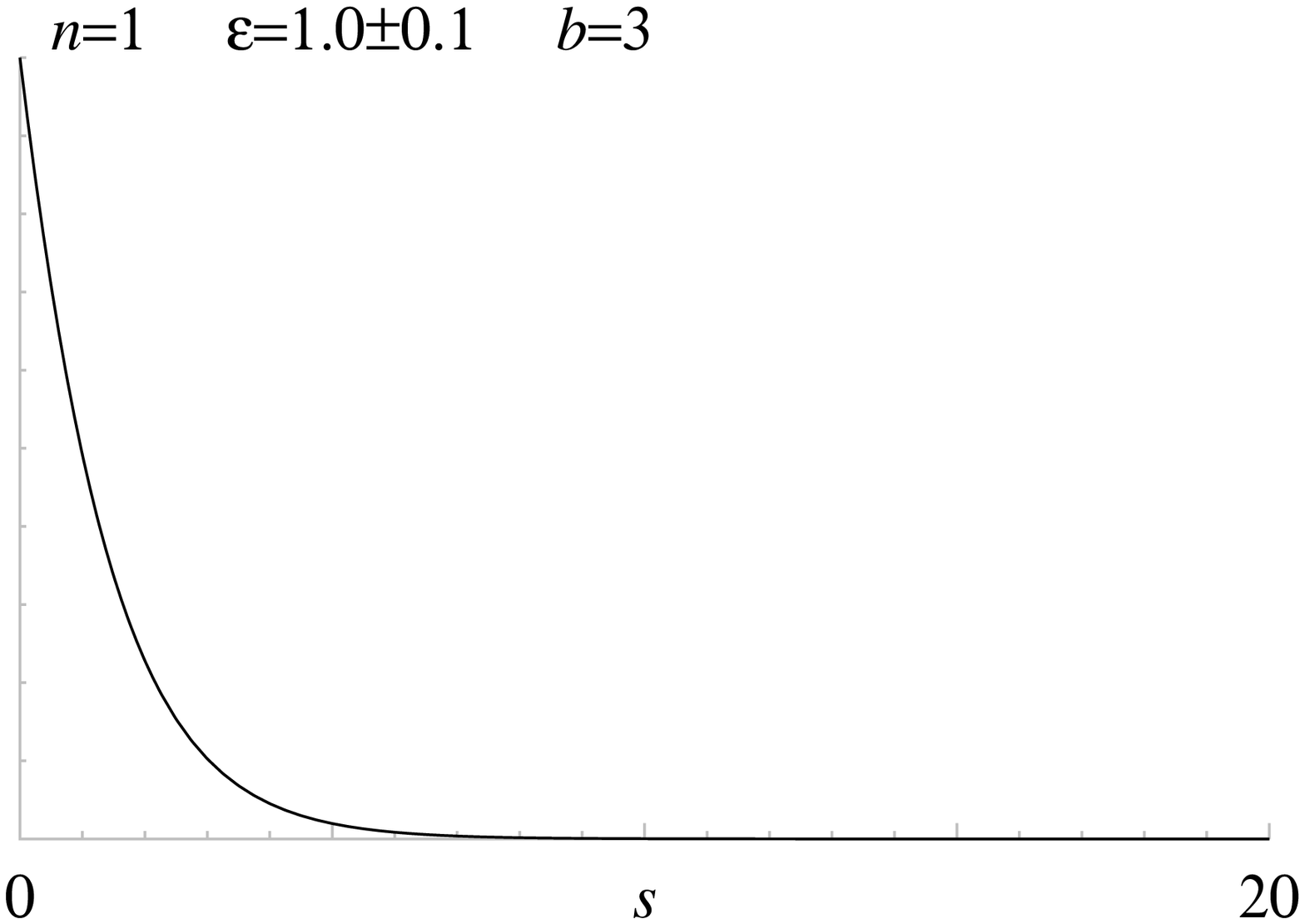}\hskip-0.2in
\includegraphics[width=3.15in]{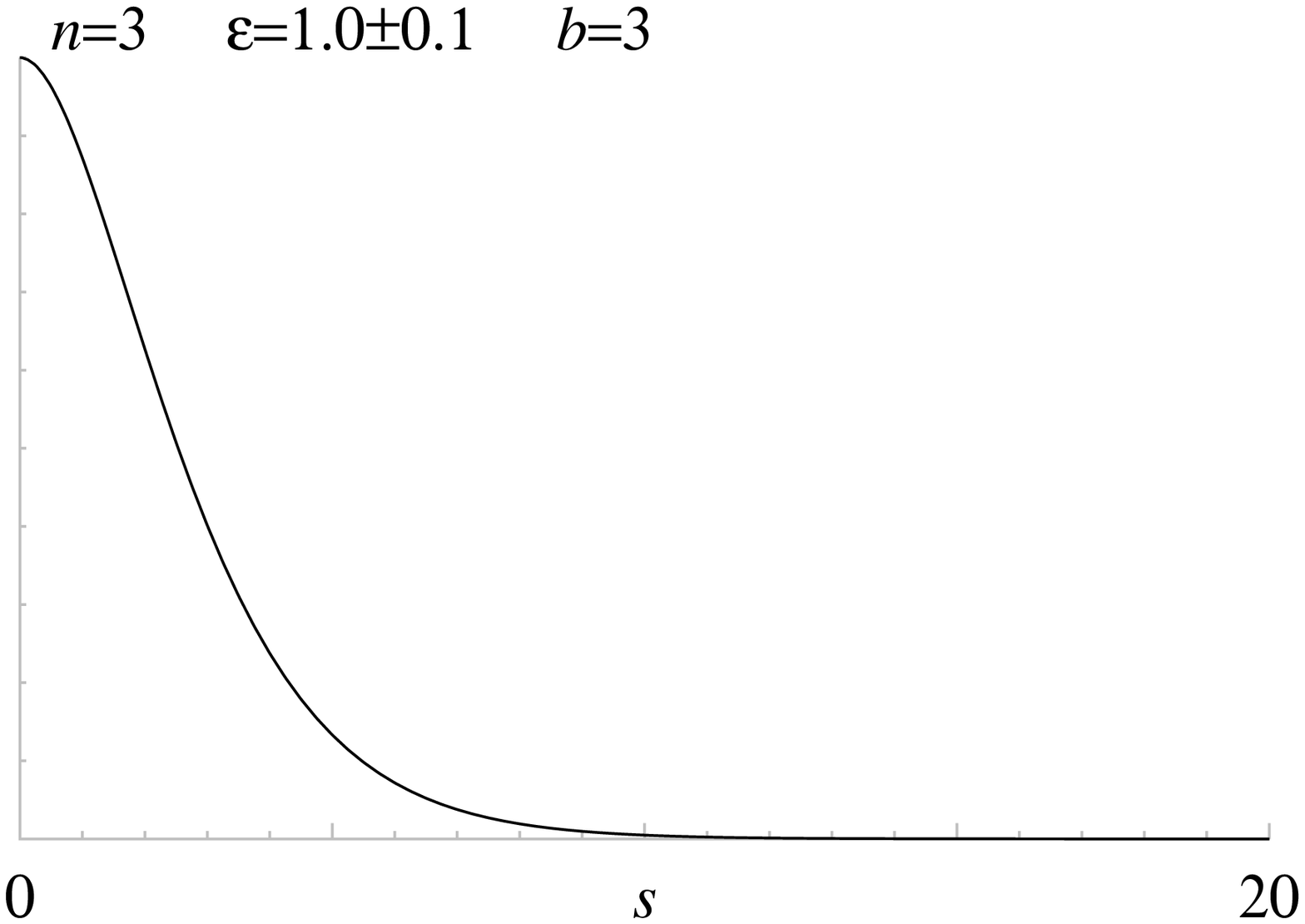}\\
\includegraphics[width=3.15in]{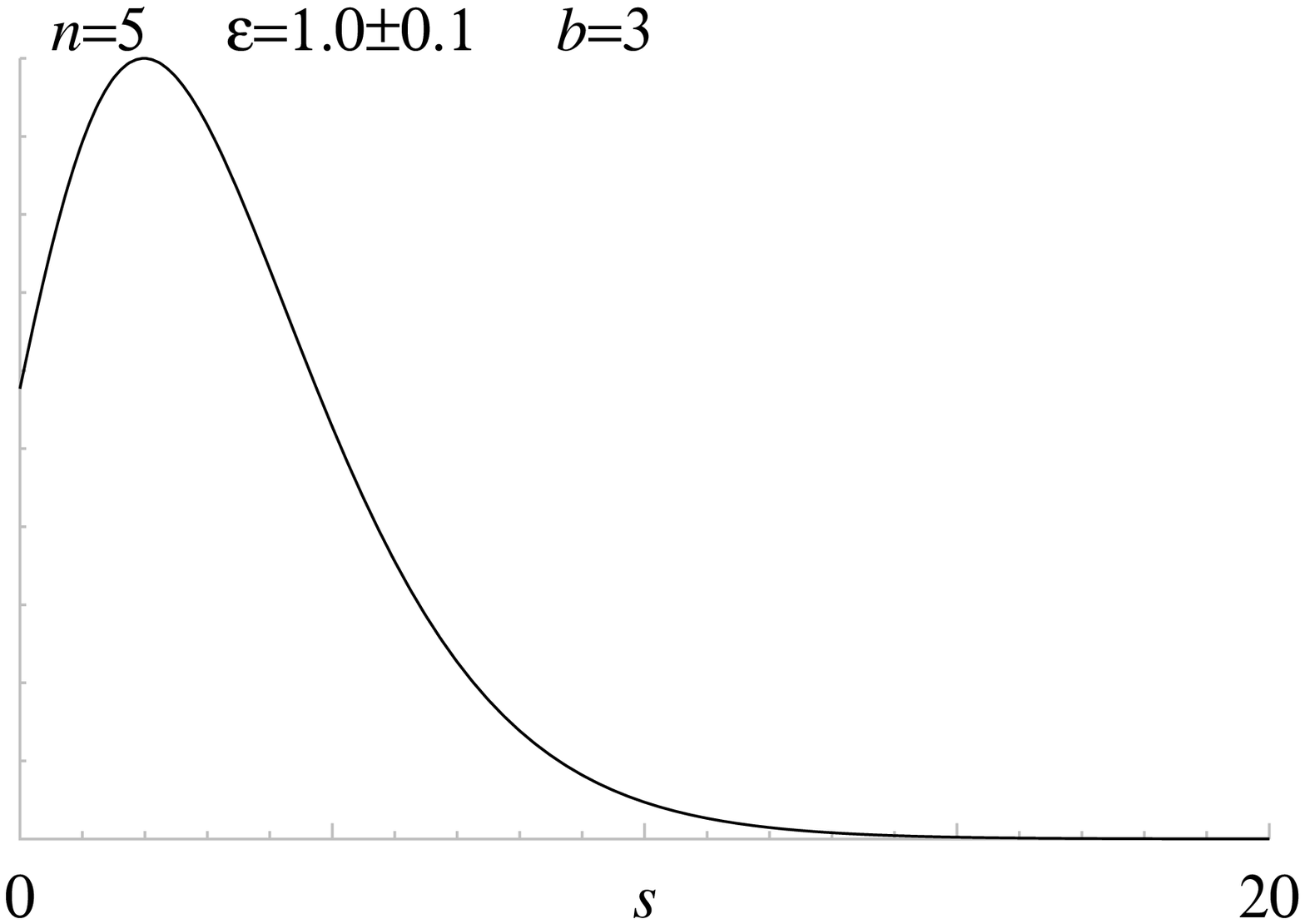}\hskip-0.2in
\includegraphics[width=3.15in]{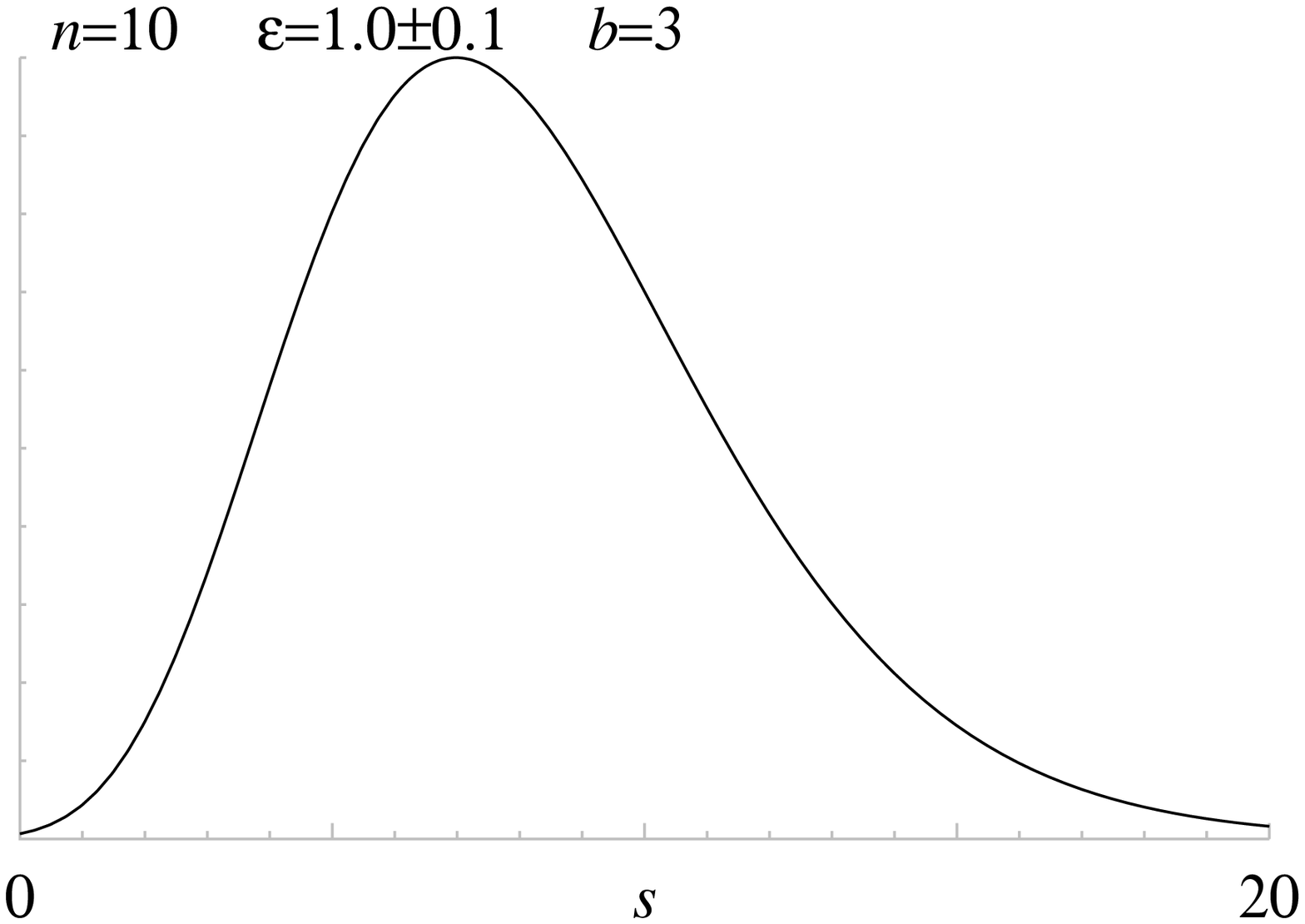}\\
\caption{Posterior densities $p(s|b,n)$ vs $s$ for $n=1$, 3, 5, 10.
In each case, $b=3$ and $\epsilon=1.0\pm0.1$ (i.e.\ $\kappa=100$ and $m$=99).}
\label{pdfs}
\end{center}
\end{figure}

Figure~\ref{pdfs} illustrates the shape of the posterior for $s$
(i.e.\ marginalized over $\epsilon$) in the case of a nominal 10\%
uncertainty on $\epsilon$, and an expectation of 3 background
events. Plots are shown for 1, 3, 5, and 10 observed events.  The
posterior evolves gracefully from being strongly peaked at $s=0$ to a
roughly Gaussian shape that excludes the neighborhood near $s=0$ with
high probability. Technically, the posterior would be described as a
mixture of $n+1$ Beta distributions of the 2nd kind\footnote{The 2nd Beta
distribution is also known as ``$\mathrm{Beta}'$''
(i.e.\ ``Beta prime''), ``inverted Beta'', ``Pearson Type~VI'',
``Variance-Ratio'', ``Gamma-Gamma'', ``F'', ``Snedecor'',
``Fisher-Snedecor''\ldots.}, giving
it a tail at high $s$ that is heavier than that of a Gaussian.

\subsection{Upper limits}

In this note, our main goal is to obtain a
Bayesian upper limit $s_\mathrm{u}$ from our observation of $n$
events.  It is by integrating the posterior p.d.f.\ out to
$s=s_\mathrm{u}$ that an upper limit is calculated: a $\beta=90\%$
upper limit is defined so that the integral of the posterior from
$s=0$ to $s=s_\mathrm{u}$ is 0.9.  The probability (in the Bayesian
sense) of $\st<s_\mathrm{u}$ is then exactly $\beta$.

Table~\ref{ltablex} shows the upper limits ($\beta=0.9$) for $n=0$--20
observed events with $b=0$--8 and $\epsilon=1.0\pm0.1$.
(Integer values of $b$ are chosen for illustration purposes only;
$b$ can, of course, take any real value $\ge0$.)

One notices
that when $n=0$, the limit is independent of the expected background
$b$. This is required in the Bayesian approach: we know that exactly
zero background events were produced (when no events at all were
produced), and this knowledge of what {\em did} happen makes what
might have happened superfluous.  An interesting corollary
is, in the case of no events observed, uncertainties in estimating
the background rate are of no consequence in the Bayesian approach,
and must not contribute any systematic uncertainty to the limit.
This reasoning does not hold in the frequentist framework,
where what might have happened definitely does influence the limit.

For comparison, limits for fixed $\epsilon=1$ with $b=0$ or $b=3$ are
also shown in Table~\ref{ltablex}. It is interesting that these two
columns start out equal at $n=0$ and differ by almost exactly 3 for
$n>11$. In contrast, the difference between the $b=0$ and $b=3$
columns for $\epsilon=1.0\pm0.1$ is already greater than 3 at $n=6$, and
continues to grow as $n$ increases; it is not clear whether
the difference approaches a finite value as $n\to\infty$.
In any case, the limits for $\epsilon=1$ exactly are all smaller
than the corresponding limits for $\epsilon=1.0\pm0.1$,
as expected.

\subsection{Coverage}

The main quantity of interest in this subsection is the frequentist
coverage probability $C$ as a function of \st\ (for fixed \et\ and
$b$).  Because both the main and the subsidiary measurements involve
observing a discrete number of events, the function $C(\st)$ will have
many discontinuities. On the other hand, $C(\et)$ will be continuous
(for fixed \st). The explanation of this effect is as follows:

\begin{quotation}
\footnotesize

The measured data are $n$ events in the main measurement and $m$ events
in the subsidiary measurement.  For each observed outcome $(n,m)$
there is a limit $s_\mathrm{u}(n,m)$. This limit includes the effect
of marginalization over $\epsilon$.

All $(n,m)$ with $n\ge0$ and $m\ge0$ are possible, and the probability
$P$ of observing $(n,m)$ can be calculated as the product of two
Poissons.  (It will depend on \st, \et, \ldots)  If we look at all the
possible limits we can obtain,
\begin{equation}
           \{ s_\mathrm{u}(n,m) | n\ge0\ \mathrm{and}\ m\ge0\}
\end{equation}
and sort them in increasing $s_\mathrm{u}$, the $s_\mathrm{u}$ are
countably infinite in number and dense in the same way that rational
numbers are dense in the reals.

To compute the coverage as a function of \st, we simply add up all
the probabilities of obtaining $(n,m)$ with $s_\mathrm{u}(n,m)\ge\st$:
\begin{equation}
C=\!\!\sum_{(n,m)\in\mathcal{A}}\!\!P(n,m)\qquad\qquad
\mathcal{A}=\{(n,m) | \st\le
s_\mathrm{u}(n,m)\ \mathrm{and}\ n\ge0\ \mathrm{and}\ m\ge0\}
\end{equation}
This sum is over a countably infinite number of terms.  If we
increase $\st$ slightly to $\st+ds$ and recalculate the coverage, we
have to drop all the terms
\begin{equation}
   \{ (n,m) | \st \le s_\mathrm{u}(n,m) \le \st+ds \}
\end{equation}
from the previous sum (the $P(n,m)$ for each term also changes
continuously with $\st$, but this is no problem).  If there are $M>0$
such terms, there are $M$ discontinuities in the coverage in the
interval $[\st,\st+ds]$, since $P(n,m)$ for each of these is finite,
and we lose them one by one as we sweep across the interval
$[\st,\st+ds]$.

But it seems that, in general, we can always find a solution to
$\st\le s_\mathrm{u}(n,m)\le\st+ds$ for finite $ds$ by going out to larger and
larger $n$ and $m$.  So, although the discontinuity may be tiny, we can
always find a finite discontinuity in any finite interval of $\st$.

On the other hand, if we keep $\st$ fixed and vary $\et$, we always
sum over the same set of $(n,m)$, since the definition of
$\mathcal{A}$ does not involve \et, and $P(n,m)$ is continuous in
\et.  So the coverage is continuous as a function of \et\ for \st\
fixed.

\end{quotation}

\begin{figure}[p]
\begin{center}
\includegraphics[width=\textwidth]{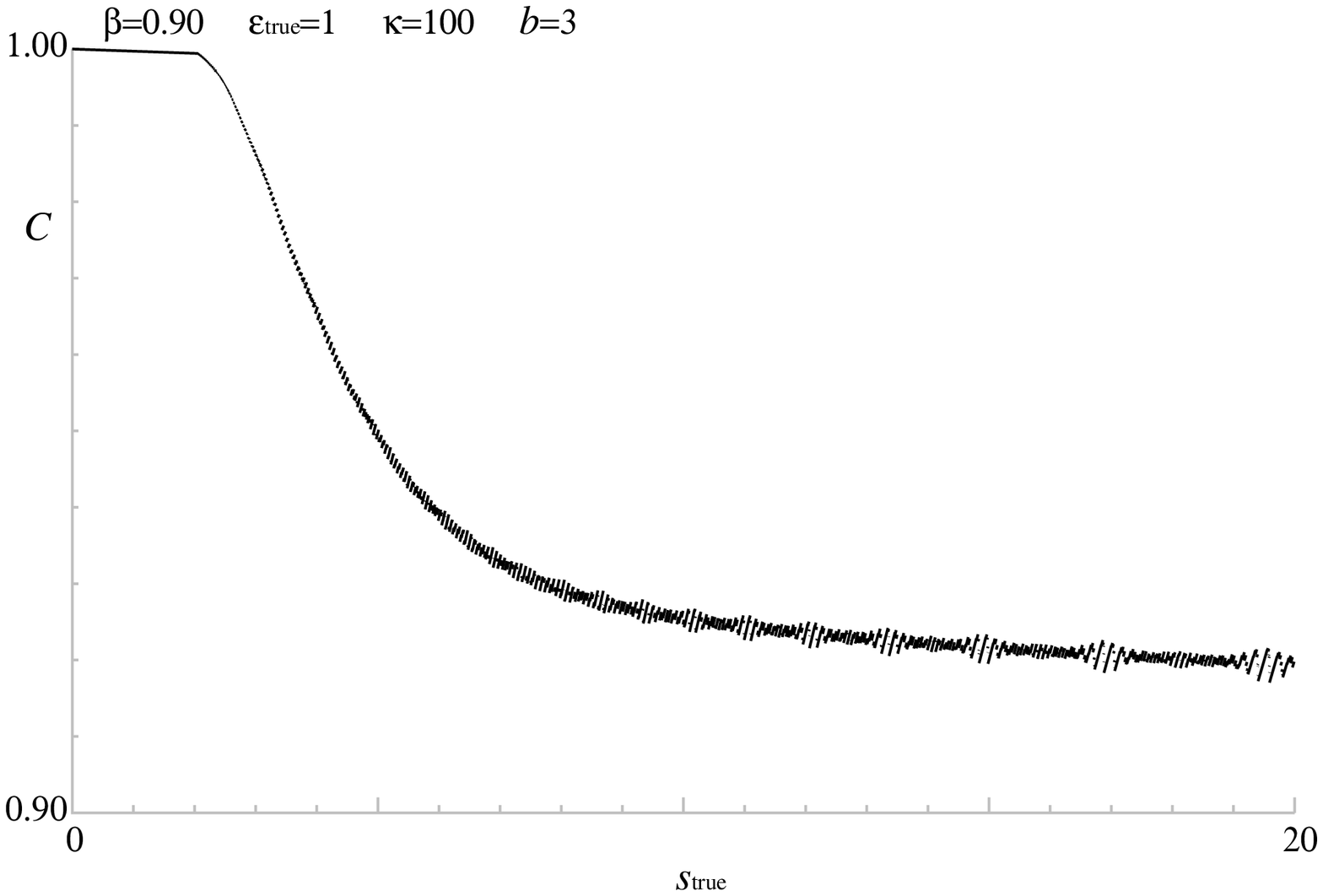}
\caption{Coverage of 90\% upper limits as a function of \st\ for
$\et=1$, nominal 10\% uncertainty of the subsidiary
measurement of $\epsilon$, and expected background $b=3$.}
\label{l100}
\end{center}
\end{figure}

\begin{figure}[p]
\begin{center}
\includegraphics[width=\textwidth]{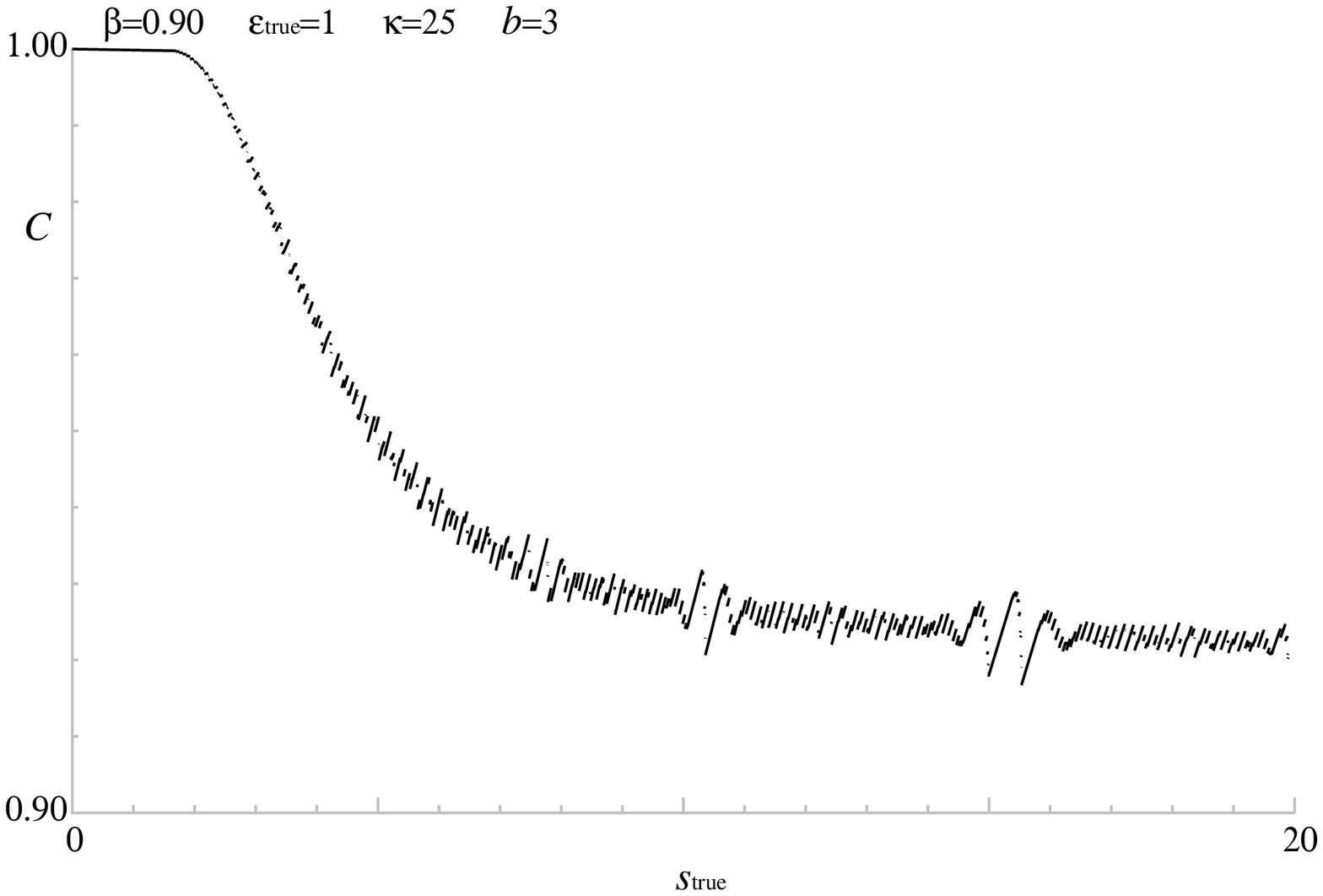}
\caption{Coverage of 90\% upper limits as a function of \st\ for
$\et=1$, nominal 20\% uncertainty of the subsidiary
measurement of $\epsilon$, and expected background $b=3$.}
\label{l25}
\end{center}
\end{figure}

Plotting a curve that is discontinuous at every point is somewhat
problematical. The solution adopted here is to plot the coverage as
straight line segments between the discontinuities, ignoring any
discontinuities with $|\Delta C|<10^{-4}$.  Figure~\ref{l100} shows
$C(\st)$ for the case $\beta=90\%$, $\et=1$, nominal 10\% uncertainty
of the subsidiary measurement of $\epsilon$, and $b=3$. We observe
that $C(\st)>\beta$ in this range, and it is not clear numerically
whether $C(\st)\to\beta$ as $\st\to\infty$. The same conclusions hold
for Fig.~\ref{l25}, which illustrates the same situation with a 20\%
nominal uncertainty for the $\epsilon$-measurement.

\begin{figure}[p]
\begin{center}
\includegraphics[width=\textwidth]{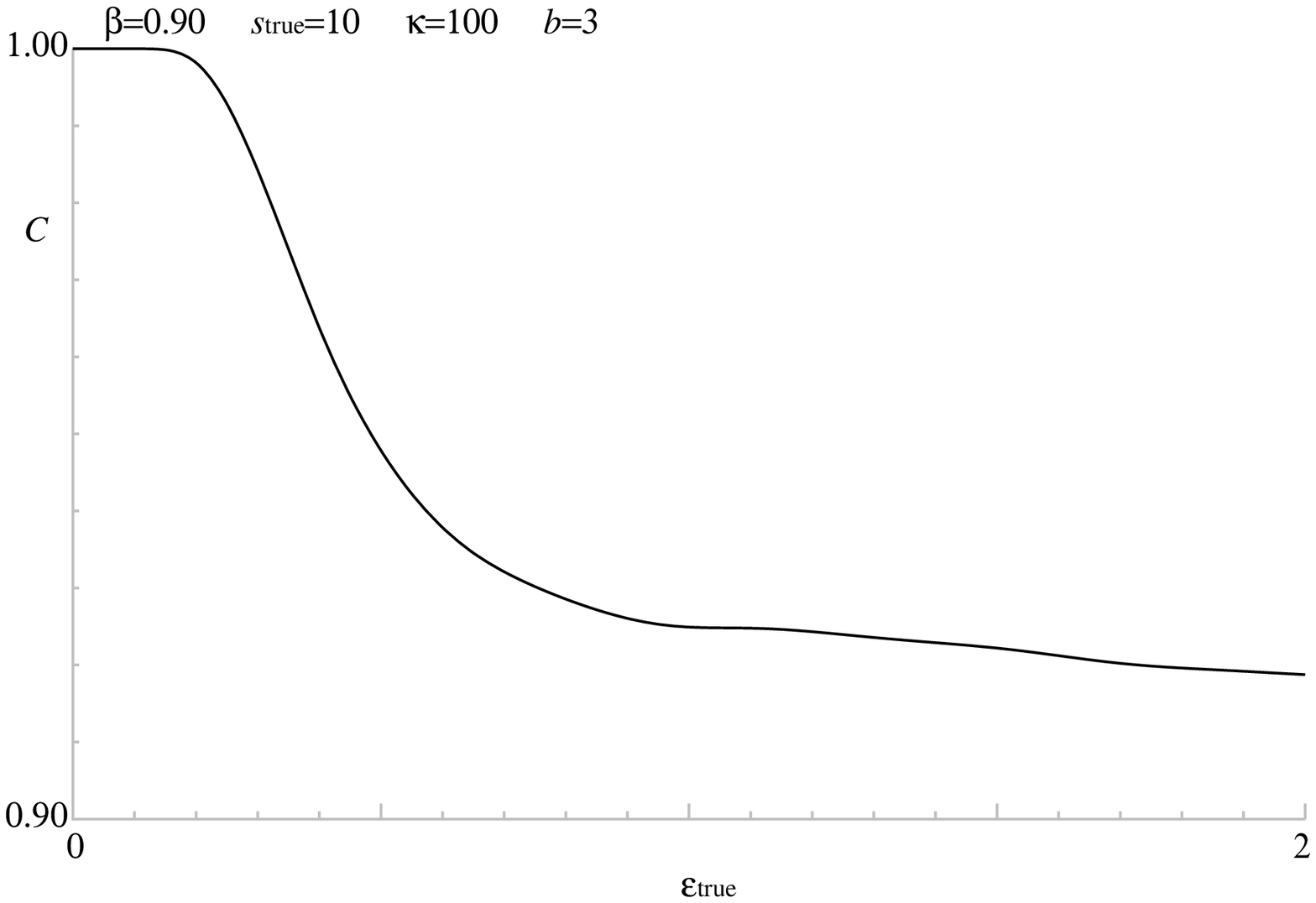}
\caption{Coverage of 90\% upper limits as a function of \et\ for
$\st=10$, nominal 10\% uncertainty of the subsidiary
measurement of $\epsilon$, and expected background $b=3$.}
\label{ecov}
\end{center}
\end{figure}

Figure~\ref{ecov} shows $C(\et)$ for $\beta=90\%$, $\st=10$,
$\kappa=100$, and $b=3$---continuous as advertised.  The shape of the
curve is quite similar to that of Figs.~\ref{l100} and \ref{l25}, so
it seems that the coverage probability (with $b$ fixed) is
approximately a function of just the product of \et\ and \st. This
approximate rule is likely to fail in the limit as $\et\to0$ and
$\st\to\infty$, for example, but it seems to hold when \et\ and
\st\ are at least of the same order of magnitude.

\begin{figure}[p]
\begin{center}
\includegraphics[width=\textwidth]{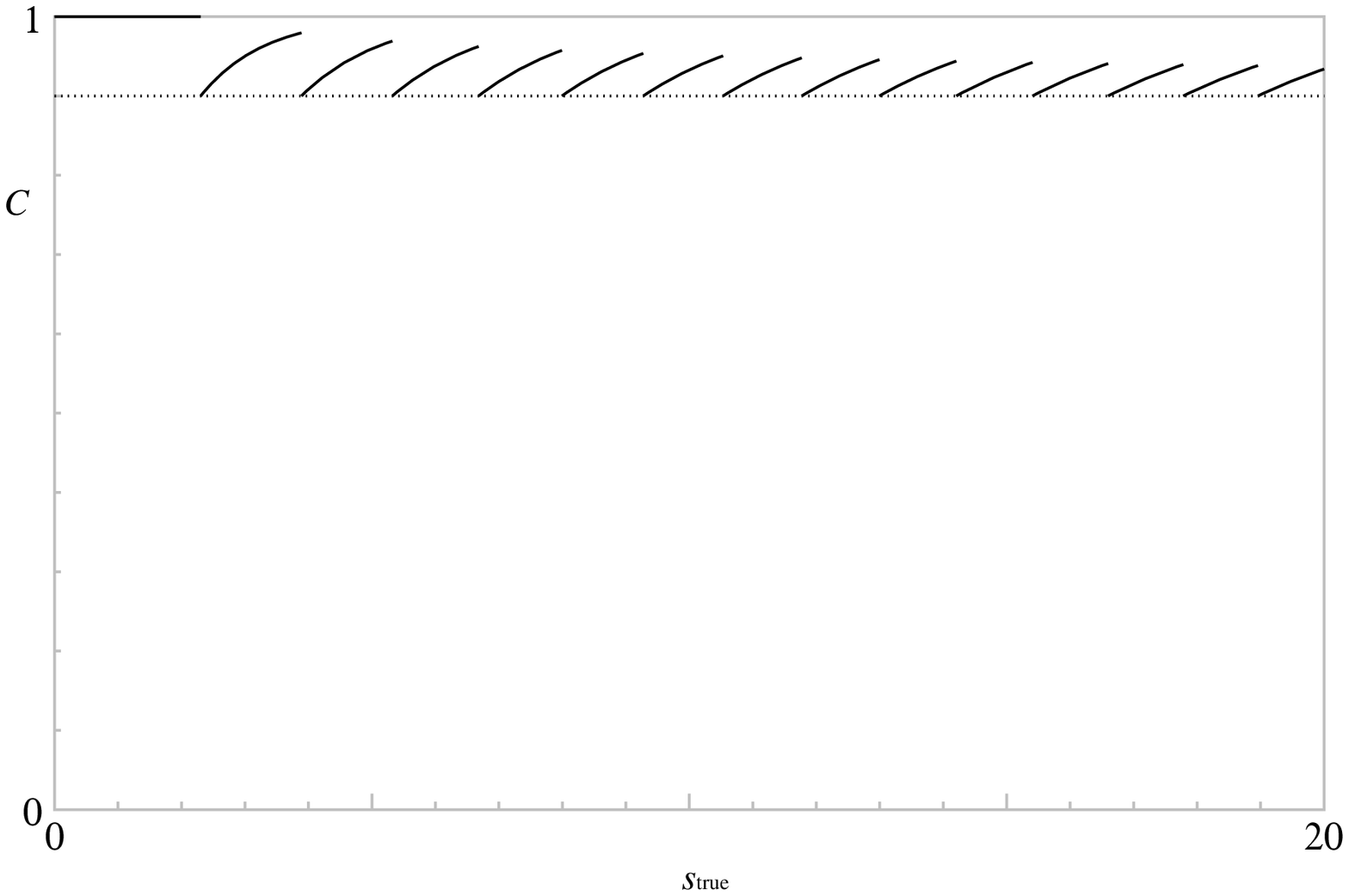}
\caption{
Coverage as a function of the true signal rate $s$ for Bayes 90\%
limits, for the simple case of no background and no uncertainty on
$\epsilon = 1$. The dotted line at $C=0.9$ is given to
show that the coverage never falls below 90\% (in this
simple case).}
\label{fig:Bayes}
\end{center}
\end{figure}

When $\et\st$ is small, of order 1 or less, the coverage is
$\sim$100\%, as in the simple case of Fig.~\ref{fig:Bayes}.
Otherwise, the behavior of coverage in Figs.~\ref{l100}--\ref{ecov}
is superior to
that of Fig.~\ref{fig:Bayes}, which has
a much larger amplitude of oscillation.

\begin{figure}[p]
\begin{center}
\includegraphics[width=\textwidth]{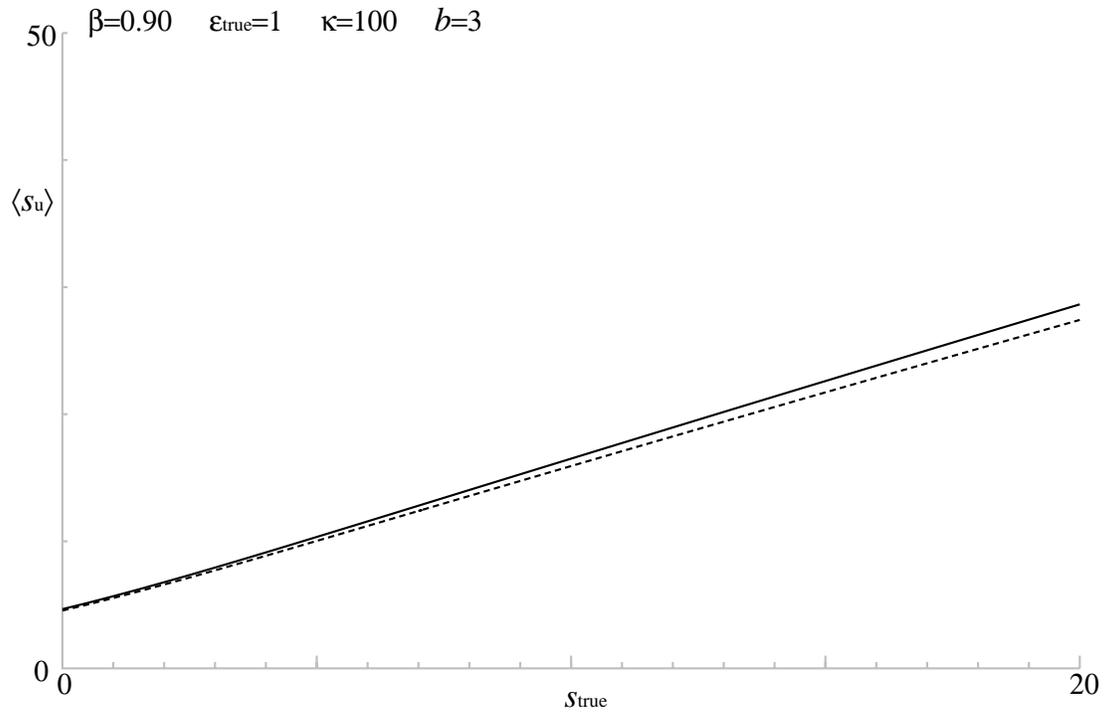}
\caption{Sensitivity of 90\% upper limits as a function of \st\ for
$\et=1$, nominal 10\% uncertainty of the subsidiary
measurement of $\epsilon$, and expected background $b=3$.
For reference, the sensitivity for $\sigma_\epsilon=0$ is also 
given (dashed).}
\label{sens}
\end{center}
\end{figure}

Another frequentist quantity that characterizes the performance of a
limit scheme is the sensitivity, defined as the mean of
$s_\mathrm{u}$.  Figure~\ref{sens} shows the sensitivity as a function
of \st\ for the case of Fig.~\ref{l100}; $\langle
s_\mathrm{u}\rangle$ is observed to be nearly linearly dependent on
\st. There is one complication here: when the subsidiary
measurement observes $m=0$ events, and the prior for $s$ is flat,
$s_\mathrm{u}=\infty$. Since the Poisson probability of obtaining
$m=0$ is always finite, $\langle s_\mathrm{u}\rangle$ is consequently
infinite. So we must exclude the $m=0$ case from the
definition of $\langle s_\mathrm{u}\rangle$. (In Fig.~\ref{sens}
the probability of obtaining $m=0$ is $e^{-100}\simeq4\times10^{-44}$.)

\subsection{Other priors for $s$}

A weakness of the Bayesian approach is that there is no
universally accepted method to obtain a unique ``non-informative''
or ``objective'' prior p.d.f. Reference~\cite{roots}, for example,
states:
\begin{quote}
Put bluntly: data cannot ever speak entirely for themselves; every
prior specification has {\em some} informative posterior or predictive
implications; and ``vague'' is itself much too vague an idea to be
useful.  There is no ``objective'' prior that represents ignorance.
\end{quote}
Nevertheless, Ref.~\cite{roots} does derive a $1/\sqrt{s}$ ``reference
prior'' for the simple Poisson case, which is claimed to have ``a
minimal effect, relative to the data, on the final inference''.  This
is to be considered a ``{\em default} option when there are
insufficient resources for detailed elicitation of actual prior
knowledge''.

Reference~\cite{ref:Jeffreys} attempts to discover the optimal form for
prior ignorance by considering the behavior of the prior under
reparameterizations.  For the case in question, the form $1/s$ clearly
has the best properties in this respect.

We are using an flat ($s^0$) prior for this study, which
seems to be the most popular choice, but the Appendix works out the
form of the posterior using an $s^{\alpha-1}$ prior, so we can briefly
here summarize the results for the $1/s$ and $1/\sqrt{s}$ cases:

The $1/s$ prior leads to an unnormalizable posterior for all observed
$n$ when $b>0$. The posterior becomes a $\delta$-function at $s=0$,
$s_\mathrm{u}=0$ for any $\beta$, and the coverage is consequently
zero for all $\st>0$. This clearly is a disaster.

The $1/\sqrt{s}$ prior results in a posterior p.d.f.\ qualitatively
similar in shape to those of Fig~\ref{pdfs}, except that the p.d.f.\
is always infinite at $s=0$. For $n\gg b$, this produces an extremely
thin ``spike'' at $s=0$, which has a negligible contribution to the
integral of the posterior p.d.f. A more significant difference (for
frequentists) between the $1/\sqrt{s}$ and the $s^0$ case is that the
coverage probability is significantly reduced: for the case of
Fig.~\ref{l100} the $1/\sqrt{s}$ prior pushes the minimum coverage
down to $\sim$0.87.  So the $1/\sqrt{s}$ prior leads to violation of
the frequentist coverage requirement; it undercovers for some
values of \st.

One might also seek to
further improve the coverage properties by adopting an intermediate
prior. For example, an $s^{-0.25}$ prior would reduce the level of
overcoverage obtained with the $s^0$ prior.  How acceptable this
approach would be within the Bayesian Statistical community is an
interesting question.

It should be noted that all the prior p.d.f.'s considered in this note
are ``improper priors''---they cannot be correctly normalized: In the
case of the $s^0$ and $1/\sqrt{s}$ priors, the integral from 0 to any
value $s_0$ is finite, while the integral from $s_0$ to infinity is
infinite. The corresponding integrals of the $1/s$ prior are infinite
on both sides for all $s_0>0$. Improper priors are dangerous but often
useful; ``improper posteriors'' are generally pathological.  Extra
care must be taken when employing improper priors to verify the
normalizability of the resulting posterior---when using a numerical
method to obtain the posterior, it is very easy to miss the fact that
its integral is infinite.

\subsection{Restrictions}

We summarize here the restrictions forced on the priors for $s$ and
$\epsilon$---see the Appendix for the analytical causes.  The
discussion below assumes $b>0$.  The prior for $s$ being of the form
$s^{\alpha-1}$, we must require $\alpha>0$, as discussed above.

As specified in this note, the prior for $\epsilon$, being taken from
the posterior from the subsidiary measurement with a flat prior, has
been given no freedom. Should the subsidiary measurement observe $m=0$
events, the posterior for $s$ is not normalizable when $\alpha\ge1$:
$s_\mathrm{u}=\infty$ when $m=0$ and $\alpha\ge1$.

This behavior is due to a well known effect: the $\epsilon$ prior
becomes $\kappa e^{-\kappa\epsilon}$ when $m=0$, which remains finite
as $\epsilon\to0$. All such cases\footnote{A Gaussian truncated at
$\epsilon=0$ is the standard example.} yield $s_\mathrm{u}=\infty$ when
$\alpha\ge1$; any positive $\alpha<1$ cuts off the posterior at large
$s$ sufficiently rapidly to render it normalizable. From this point of
view, a $1/\sqrt{s}$ prior may seem preferable, but on the other hand,
having $s_\mathrm{u}=\infty$ when $m=0$ seems intuitively reasonable.
(In general, we have $s_\mathrm{u}=\infty$ for $m\le\alpha-1$, but
$\alpha\ge2$ are not popular choices.)

There is another approach possible to the gamma prior for $\epsilon$:
we may simply specify by fiat the form of the prior as
\begin{equation}
p(\epsilon|\mu)d\epsilon =
{\kappa  (\kappa\epsilon)^{\mu-1}  e^{-\kappa\epsilon} \over\Gamma(\mu)}
d\epsilon
\end{equation}
where $\mu$ is no longer required to be an integer.  In practice, one
then might obtain $\mu$ and $\kappa$ from a subsidiary measurement
whose result is approximated by the gamma distribution.
In such cases,
one must require $\mu>\alpha$ to keep the posterior normalizable. Note
that in this form, $\mu/\kappa$ is the mean of the $\epsilon$ prior,
$(\mu-1)/\kappa$ is the mode, and $\mu/\kappa^2$ is the variance.
The subsidiary measurement is often analysed by other experimenters,
who chose statistics to quote for their central value and uncertainty
(omitting additional likelihood information).
It is then important to obtain $\mu$ and $\kappa$ in a consistent way
from the information supplied by the subsidiary measurement.  If
$\epsilon$, for example, were estimated by a maximum likelihood
method, one would identify the estimate with $(\mu-1)/\kappa$ rather
than $\mu/\kappa$.

\section{Conclusions}

Results have been presented on the performance of a purely Bayesian
approach to the issue of setting upper limits on the rate of a
process, when $n$ events have been observed in a situation where the
expected background is $b$ and where the efficiency/acceptance factor
$\epsilon \pm \sigma_{\epsilon}$ has been determined in a subsidiary
experiment.
We find that this approach, when using a flat prior for the rate,
results in modest overcoverage.
Plots of the expected sensitivity of such a measurement
and of the coverage of the upper limits are given.
It will be
interesting to compare these with the corresponding plots for other
methods of extracting upper limits, to be given in future notes.
Reference~\cite{software} provides the limit calculating software
associated with this study in the form of C~functions.

\section{Appendix A---Analytical Details}


Here we present the details of the analytical calculation of the
posterior p.d.f.\ for $s$. For generality, we work through the
calculation with a $s^{\alpha-1}$ prior; a flat prior is then the
special case $\alpha=1$.

\subsection{Posterior for $s$ with $\epsilon$ and $b$ fixed }

We measure $n$ events from a process with Poisson rate $\epsilon s+b$,
and we want the Bayesian posterior for $s$, given improper prior
$s^{\alpha-1}$.  We compute the posterior for fixed $\epsilon$ and $b$
in this subsection; the calculation with our prior
for $\epsilon$ follows in the next subsection. We have
\[
\mbox{posterior:}\quad p(s|\epsilon,b,n)ds={1\over\mathcal{N}_s}
e^{-\epsilon s}(\epsilon s+b)^ns^{\alpha-1}ds
\]
where all factors not depending on $s$ have already been absorbed
into the normalization constant $\mathcal{N}_s$, which is defined by
\[
\mathcal{N}_s=
\int_0^\infty e^{-\epsilon s}(\epsilon s+b)^ns^{\alpha-1}ds=
{b^{n+\alpha}\over\epsilon^\alpha}
\int_0^\infty e^{-bu}u^{\alpha-1}(1+u)^ndu\qquad(u=s\epsilon/b)
\]
where we have performed the indicated change of variable.

Expanding
$(1+u)^n$ in powers of $u$ using the binomial theorem, we get
\[
(1+u)^n=n!\sum_{k=0}^n{u^{n-k}\over(n-k)!k!}\qquad\Rightarrow\qquad
\mathcal{N}_s=
n!\epsilon^{-\alpha}
\sum_{k=0}^n{\Gamma(\alpha+n-k)b^k\over(n-k)!k!}
\]
Recognizing this as of the general hypergeometric form, we write it as
\[
\mathcal{N}_s=
\epsilon^{-\alpha}
\Gamma(\alpha+n)\left[1+{n\over\alpha+n-1}{b\over1!}+
{n(n-1)\over(\alpha+n-1)(\alpha+n-2)}{b^2\over2!}+\cdots\right]
\]
to make the hypergeometric nature more explicit.  Using the modern
notation\cite{ff} for the falling factorial
\[
z^{\underline{k}}\equiv
{\Gamma(z+1)\over\Gamma(z-k+1)}=z(z-1)(z-2)\cdots(z-k+1)
\]
this is expressed as
\[
\mathcal{N}_s=
\epsilon^{-\alpha}\Gamma(\alpha+n)
\sum_{k=0}^n{n^{\underline{k}}\over(\alpha+n-1)^{\underline{k}}}{b^k\over k!}=
\epsilon^{-\alpha}\Gamma(\alpha+n)M(-n,1-n-\alpha,b)
\]
where $M$ is the notation of \cite{as}.
($M$, a confluent hypergeometric function, is often written $_1F_1$,
and the relation given here is only valid for integer $n\ge0$.)  Note
that $M(-n,1-n-\alpha,b)$ is a polynomial of order $n$ in $b$ (for $n$
a non-negative integer), and is related to the Laguerre polynomials.
When $\alpha=1$, we get $M(-n,-n,b)$, which is related to the Incomplete
Gamma Function.
When $\alpha=0$, we get $M(-n,1-n,b)$, which is infinite, so we require
that $\alpha>0$.

Our posterior probability density for fixed $\epsilon$
is then given by
\[
p(s|\epsilon,b,n)ds=
{\epsilon^\alpha e^{-\epsilon s}(\epsilon s+b)^ns^{\alpha-1}
\over\Gamma(\alpha+n)M(-n,1-n-\alpha,b)}ds
\]

\subsection{Posterior for $\epsilon$ of the subsidiary measurement}

The subsidiary measurement observes an integer number of events $m$,
Poisson distributed as:
\[
P(m|\epsilon) = {e^{-\kappa\epsilon}  (\kappa\epsilon)^m\over m!}
\]
where $\kappa$ is a real number (connecting the subsidiary measurement to the
main measurement) whose uncertainty is negligible, so $\kappa$ can safely be
treated as a fixed constant.  $\kappa$ might be thought of, for example, as
based on a cross section that is exactly calculable by theory.  There
is negligible (i.e.\ zero) background in the subsidiary measurement.

The prior for $\epsilon$ is specified to be flat.
The Bayesian posterior p.d.f.\ for $\epsilon$ is then
\[
p(\epsilon|m) = {\kappa  (\kappa\epsilon)^m  e^{-\kappa\epsilon} \over m!}
\]
(or $\Gamma(m+1)$ instead of $m!$ in the denominator if you prefer).
This is known as a gamma distribution.

The mean and rms of this posterior p.d.f.\ summarize the result of the
subsidiary measurement as:
\[
\epsilon = {m + 1\over\kappa}   \pm
{\sqrt{m + 1}\over\kappa} = \epsilon_0 \pm \sigma_\epsilon 
\]
Note that the observed data quantity in the subsidiary measurement is
an integer $m$, while the quantity being measured by the subsidiary
measurement is a positive real number $\epsilon$.

\subsection{Posterior for $s$ with gamma prior for $\epsilon$ ($b$ fixed)\label{postpdf}}

Next we compute the joint posterior $p(s,\epsilon|b,n)dsd\epsilon$
using the $s^{\alpha-1}$ prior for $s$ and
our gamma distribution prior (i.e.\ the posterior derived
from the subsidiary measurement) for $\epsilon\ge0$
\[
\mbox{prior for $\epsilon$:}\quad \pi(\epsilon)d\epsilon=
{(\kappa\epsilon)^\mu e^{-\kappa\epsilon}\over\Gamma(\mu)}
{d\epsilon\over\epsilon}
\qquad\qquad\mu=m+1=(\epsilon_0/\sigma_\epsilon)^2\qquad
\kappa=\epsilon_0/{\sigma_\epsilon}^2
\]
where it is convenient to write $\mu$ for $m+1$.
We have for the joint posterior p.d.f.
\[
p(s,\epsilon|b,n)dsd\epsilon={1\over\mathcal{N}_{s,\epsilon}}
\pi(\epsilon)e^{-\epsilon s}(\epsilon s+b)^ns^{\alpha-1}dsd\epsilon
\]
where
\[
\mathcal{N}_{s,\epsilon}=\int_0^\infty\!\!\!\int_0^\infty\!\!
\pi(\epsilon)e^{-\epsilon s}(\epsilon s+b)^ns^{\alpha-1}dsd\epsilon=
\int_0^\infty\!\!\pi(\epsilon)\mathcal{N}_sd\epsilon
\]
We calculated $\mathcal{N}_s$ above, so we have
\[
\mathcal{N}_{s,\epsilon}=
\Gamma(\alpha+n)M(-n,1-n-\alpha,b)
\int_0^\infty\!\!\epsilon^{-\alpha}\pi(\epsilon)d\epsilon
\]
\[
\mathcal{N}_{s,\epsilon}=
\kappa^\alpha\Gamma(\mu-\alpha)\Gamma(\alpha+n)M(-n,1-n-\alpha,b)/\Gamma(\mu)
\]
\[
p(s,\epsilon|b,n)dsd\epsilon=
{\kappa^{\mu-\alpha}\epsilon^{\mu-1}s^{\alpha-1}(\epsilon s+b)^n
e^{-(s+\kappa)\epsilon}\over
\Gamma(\mu-\alpha)\Gamma(\alpha+n)M(-n,1-n-\alpha,b)}dsd\epsilon
\]
The marginalized posterior for $s$ can then be expressed as
\[
p(s|b,n)ds=\left[\int_0^\infty\!\!p(s,\epsilon|b,n)d\epsilon\right]ds=
{s^{\alpha-1}\kappa^{\mu-\alpha}\mathcal{I}_\epsilon\over
\Gamma(\mu-\alpha)\Gamma(\alpha+n)M(-n,1-n-\alpha,b)}ds
\]
where the integral $\mathcal{I}_\epsilon$ is given by
\[
\mathcal{I}_\epsilon=
\int_0^\infty\epsilon^{\mu-1}e^{-(s+\kappa)\epsilon}
(\epsilon s+b)^nd\epsilon
\]
The same procedure that was used for the normalization integral can
be applied here, producing
\[
\mathcal{I}_\epsilon=
{b^{\mu+n}\over s^\mu}
\int_0^\infty u^{\mu-1}e^{-b(1+\kappa/s)u}(1+u)^ndu
\]
\[
\mathcal{I}_\epsilon=
{s^nn!\over(s+\kappa)^{\mu+n}}
\sum_{k=0}^n{\Gamma(\mu+n-k)\over
(n-k)!k!}\left[b(s+\kappa)\over s\right]^k
\]
\[
\mathcal{I}_\epsilon=
{s^n\over(s+\kappa)^{\mu+n}}\Gamma(\mu+n)
M(-n,1-n-\mu,b(s+\kappa)/s)
\]
\[
p(s|b,n)ds=
{\Gamma(\mu+n)\over\Gamma(\mu-\alpha)\Gamma(\alpha+n)}
{s^{\alpha+n-1}\kappa^{\mu-\alpha}\over(s+\kappa)^{\mu+n}}
{M(-n,1-n-\mu,b(s+\kappa)/s)\over M(-n,1-n-\alpha,b)}ds
\]
which has a particularly simple form when the background term is zero:
\[
p(s|b=0,n)ds=
{\Gamma(\mu+n)\over\Gamma(\mu-\alpha)\Gamma(\alpha+n)}
{s^{\alpha+n-1}\kappa^{\mu-\alpha}\over(s+\kappa)^{\mu+n}}ds
\]
a Beta distribution of the 2nd kind. Note that we must require
$\mu>\alpha>0$ to obtain a normalizable posterior.

Our posterior p.d.f.\ for $s$ with $\epsilon$ (and $b$) fixed is
recovered exactly by taking the limit of $p(s|b,n)$ as
$\sigma_\epsilon\to0$. This means that the limit of $s_{\mathrm{u}}$
as $\sigma_\epsilon\to0$ is identical to the value of $s_{\mathrm{u}}$
when $\epsilon$ is known exactly. This property may seem obvious, but
it is violated by some frequentist methods of setting limits,
so it is worth mentioning.

\subsection{Calculating the limit \label{intpostpdf}}

We need to integrate $p(s|b,n)$ up to some limit $s_\mathrm{u}$, which
can be done analytically as follows.
\[
\int_0^{s_\mathrm{u}}\!\!p(s|b,n)ds=
{\Gamma(\mu+n)\over\Gamma(\mu-\alpha)\Gamma(\alpha+n)}
\int_0^{s_\mathrm{u}\over s_\mathrm{u}+\kappa}
t^{\alpha+n-1}(1-t)^{\mu-\alpha-1}
{M(-n,1-n-\mu,b/t)\over M(-n,1-n-\alpha,b)}dt
\]
where the substitution $t={s\over s+\kappa}$ has been performed.
Re-expanding the polynomial $M$ and integrating term by term yields
\[
\int_0^{s_\mathrm{u}}\!\!p(s|b,n)ds=
\sum_{k=0}^n
{I_x(\alpha+n-k,\mu-\alpha)n^{\underline{k}}\over(\alpha+n-1)^{\underline{k}}}
{b^k\over k!}\Bigg/\sum_{k=0}^n
{n^{\underline{k}}\over(\alpha+n-1)^{\underline{k}}}
{b^k\over k!}\qquad\left(x={s_\mathrm{u}\over s_\mathrm{u}+\kappa}\right)
\]
where $I_x$ is the standard notation for the Incomplete Beta Function
\[
I_x(q,r)\equiv
{\Gamma(q+r)\over\Gamma(q)\Gamma(r)}
\int_0^xt^{q-1}(1-t)^{r-1}dt
\]
which also satisfies the following recursion:
\[
I_x(q,r)={\Gamma(q+r)\over\Gamma(q+1)\Gamma(r)}x^q(1-x)^r+I_x(q+1,r)
\]

\subsection{Integer moments of the marginalized posterior}
Using the same technique as above, we can calculate the $j$th moment
of the posterior p.d.f.\ as 
\[
\langle s^j\rangle=\int_0^\infty\!\!s^jp(s|b,n)ds=
{(\alpha+n)^{\overline{j}}\kappa^j\over(\mu-\alpha-1)^{\underline{j}}}
{M(-n,1-n-\alpha-j,b)\over M(-n,1-n-\alpha,b)}
\]
where we utilize the rising factorial notation\cite{ff}
\[
z^{\overline{k}}\equiv
{\Gamma(z+k)\over\Gamma(z)}=z(z+1)(z+2)\cdots(z+k-1)
\]

The expression for the mean of the posterior when $\alpha=1$
can be simplified using the identity
\[
M(-n,-n-1,b)=\left(1-{b\over n+1}\right)M(-n,-n,b)+{b^{n+1}\over(n+1)!}
\]
obtaining
\[
\mathrm{mean}(\alpha=1)=\langle s\rangle|_{\alpha=1}=
{\kappa(n+1-b)\over\mu-2}+
{\kappa b^{n+1}\over(\mu-2)n!M(-n,-n,b)}
\]
Note that the 2nd term is very small when $n\gg b$.

The recurrence relation\cite{as}
\[
r(r-1)M(q,r-1,z)+r(1-r-z)M(q,r,z)+z(r-q)M(q,r+1,z)=0
\]
leads to a recurrence relation between moments
\[
\langle s^j\rangle=
{\kappa(\alpha+n+j-1-b)\over\mu-\alpha-j}\langle s^{j-1}\rangle+
{\kappa^2b(\alpha+j-2)\over(\mu-\alpha-j+1)(\mu-\alpha-j)}
\langle s^{j-2}\rangle
\]
The special case $\alpha=1$ then yields
\[
\langle s^2\rangle|_{\alpha=1}=
{\kappa^2\over(\mu-2)(\mu-3)}\left[
(2+n-b)(1+n-b)+b+{(2+n-b)b^{n+1}\over n!M(-n,-n,b)}\right]
\]
which leads to this approximation for the variance of the posterior
\[
\mathrm{variance}(\alpha=1)\simeq
{\kappa^2(1+n)\over(\mu-2)(\mu-3)}+
{\kappa^2(1+n-b)^2\over(\mu-2)^2(\mu-3)}\qquad\qquad(n\gg b)
\]

\subsection{Posterior for $s$ with gamma priors for $\epsilon$ and $b$}

Here we very briefly consider the case where the background parameter
$b$ also acquires an uncertainty. This case is more general than the
fixed $b$ case that is the main subject of this note: the fixed $b$ case
will be the subject of additional studies employing various popular
frequentist techniques, with the goal of comparing their performance.
We judge the more general case considered in this subsection to be
more complicated than necessary for the purpose of comparing the
various methods, but it is instructive to document the fact that the
Bayesian method can easily handle the more general case.

We assume a 2nd subsidiary measurement observing $r$ events (Poisson,
as was the case for $\epsilon$), which, when combined with a flat
prior for $b$, results in a gamma posterior for $b$ of the form
\[
p(b|r)db = {\omega  (\omega b)^r  e^{-\omega b} \over r!}db
\]
where $\omega$ is a calibration constant (analogous to $\kappa$ in the
subsidiary measurement for~$\epsilon$).

The posterior for $b$ becomes the prior for $b$ in the measurement of
$s$. After determining the joint posterior $p(s,\epsilon,b|n)$ by
using our priors for $s$, $\epsilon$ and $b$, we marginalize with
respect to $\epsilon$ and $b$, resulting in
\[
p(s|n)ds=
{\Gamma(\mu+n)\over\Gamma(\mu-\alpha)\Gamma(\alpha+n)}
{s^{\alpha+n-1}\kappa^{\mu-\alpha}\over(s+\kappa)^{\mu+n}}
{F(-n,\rho;1-n-\mu;(s+\kappa)/(s\omega))\over F(-n,\rho;1-n-\alpha;1/\omega)}ds
\]
where we write $\rho=r+1$ for convenience, and $F$ is the
hypergeometric function\cite{as2}. As long as $n$ is a non-negative
integer and $\alpha>0$, $F(-n,\rho;1-n-\alpha;x)$ is a polynomial of
order $n$ in $x$ (closely related to Jacobi polynomials).

This marginalized posterior for $s$ can then be integrated, with the result
\[
\int_0^{s_\mathrm{u}}\!\!p(s|n)ds=
\sum_{k=0}^n
{I_x(\alpha+n-k,\mu-\alpha)n^{\underline{k}}\rho^{\overline{k}}
\over(\alpha+n-1)^{\underline{k}}}
{\omega^{-k}\over k!}\Bigg/\sum_{k=0}^n
{n^{\underline{k}}\rho^{\overline{k}}\over(\alpha+n-1)^{\underline{k}}}
{\omega^{-k}\over k!}\quad
\left(x={s_\mathrm{u}\over s_\mathrm{u}+\kappa}\right)
\]

These two equations closely resemble the main results of sections
\ref{postpdf} and \ref{intpostpdf}: to recover the fixed $b$
results, simply substitute $b\omega$ for $\rho$ above,
and take the limit $\omega\to\infty$.

\section{Appendix B---Average Coverage Theorem\label{ac}}

In this appendix we prove that Bayesian credible intervals have average frequentist
coverage, where the average is calculated with respect to the prior density.
We start from the Bayesian posterior density:
\begin{equation}
p(s\,|\,n)\;=\;\frac{P(n\,|\,s)\,\pi(s)}{\int_{0}^{\infty}\!P(n\,|\,s)\,\pi(s)\,ds}.
\end{equation}
For a given observed value of $n$, a credibility-$\beta$ Bayesian interval
for $s$ is any interval $[s_\mathrm{L}(n), s_\mathrm{U}(n)]$ that encloses a fraction $\beta$
of the total area under the posterior density.  Such an interval must therefore
satisfy:
\begin{equation}
\beta \;=\; \int_{s_\mathrm{L}(n)}^{s_\mathrm{U}(n)}\! p(s\,|\,n)\,ds,
\end{equation}
or, using the definition of the posterior density:
\begin{equation}
\int_{s_\mathrm{L}(n)}^{s_\mathrm{U}(n)}\! P(n\,|\,s)\,\pi(s)\,ds\;=\;\beta\;\int_{0}^{\infty}
\! P(n\,|\,s)\,\pi(s)\,ds.
\label{eq:acbci1}
\end{equation}
Now for coverage.  Given a true value $s_\mathrm{t}$ of $s$, the coverage $C(s_\mathrm{t})$ of
$[s_\mathrm{L}(n),s_\mathrm{U}(n)]$ is the frequentist probability that $s_\mathrm{t}$ is
included in that interval.  We can write this as:
\begin{equation}
C(s_\mathrm{t})\;=\; \sum_{\substack{\text{all $n$ such that:}\\[1mm]
                            s_\mathrm{L}(n)\le s\le s_\mathrm{U}(n)}}   P(n\,|\,s_\mathrm{t}).
\label{eq:acbci2}
\end{equation}
Next we calculate the average coverage $\overline{C}$, weighted by the prior $\pi(s)$:
\addtocounter{footnote}{1}
\protect\footnotetext{The best way to understand this step is to draw a diagram of
$s$ versus $n$: one is integrating and summing over the area between the curves
$s_\mathrm{L}(n)$ and $s_\mathrm{U}(n)$.  The limits on the sum and integral depend on the order
in which one does these operations and can be derived from the diagram.}
\addtocounter{footnote}{-1}
\begin{align*}
\overline{C} & \;=\;  \int_{0}^{\infty}\! C(s)\,\pi(s)\,ds ,  && \displaybreak[0]\\[6mm]
        & \;=\;  \int_{0}^{\infty}\sum_{\substack{\text{all $n$ such that:}\\[1mm]
                                                  s_\mathrm{L}(n)\le s\le s_\mathrm{U}(n)}}
                 P(n\,|\,s)\,\pi(s)\, ds,
              && \text{using equation (\protect\ref{eq:acbci2}),} \displaybreak[0]\\[6mm]
        & \;=\;  \sum_{n=0}^{\infty}\;\int_{s_\mathrm{L}(n)}^{s_\mathrm{U}(n)}\! P(n\,|\,s)\,\pi(s)\,ds ,
              && \text{interchanging integral and sum,\protect\footnotemark}\displaybreak[0]\\[6mm]
        & \;=\;  \beta\; \sum_{n=0}^{\infty}\;\int_{0}^{\infty}\! P(n\,|\,s)\,\pi(s)\,ds ,
              && \text{using equation (\protect\ref{eq:acbci1}),} \displaybreak[0]\\[6mm]
        & \;=\;  \beta\; \int_{0}^{\infty}\sum_{n=0}^{\infty} P(n\,|\,s)\,\pi(s)\,ds ,
              && \text{interchanging sum and integral,}\displaybreak[0]\\[6mm]
        & \;=\;  \beta\; \int_{0}^{\infty}\!\pi(s)\, ds ,
              && \text{by the normalization of }P(n\,|\,s), \displaybreak[0]\\[6mm]
        & \;=\;  \beta ,
              && \text{by the normalization of }\pi(s) .
\end{align*}
This completes the proof.  We have assumed here that the prior $\pi(s)$ is
proper and normalized to 1, but the proof can be generalized to improper priors
such as those we considered in this note.  A constant prior for example, can be
regarded as the limit for $s_{\max}\rightarrow\infty$ of the proper prior:
\begin{equation}
\pi(s\,|\,s_{\max}) \;=\; \frac{\vartheta(s_{\max}-s)}{s_{\max}},
\end{equation}
where $\vartheta(x)$ is $0$ if $x<0$ and $1$ otherwise.  We then {\em define}
the average coverage for the constant prior as the limit:
\begin{equation}
\overline{C}\;=\;\lim_{s_{\max}\rightarrow\, +\infty} \;
                 \int_{0}^{\infty}\! C(s)\,\pi(s\,|\,s_{\max})\,ds.
\end{equation}
The previous proof can now be applied to the argument of the limit and leads
to the same result.

The average coverage theorem remains valid when $s$ is multidimensional,
for example when it consists of a parameter of interest and one or more
nuisance parameters.  In that case one needs to average the coverage over
{\em all} the parameters.

\end{document}